\documentclass[journal]{IEEEtran}
\usepackage{xcolor,soul,framed} 
\colorlet{shadecolor}{yellow}
\usepackage[pdftex]{graphicx}
\graphicspath{{../pdf/}{../jpeg/}}
\DeclareGraphicsExtensions{.pdf,.jpeg,.png}
\usepackage[cmex10]{amsmath}
\usepackage{array}
\usepackage{algorithmic}
\usepackage{algorithm}
\usepackage{amsfonts}
\usepackage{amsmath}
\usepackage{bm}
\usepackage{bbm,dsfont}
\usepackage{booktabs}
\usepackage{blindtext}
\usepackage[colorinlistoftodos]{todonotes}
\usepackage{xcolor}
\usepackage{comment}
\usepackage{amsthm}
\usepackage{cite}
\usepackage{subfigure}
\usepackage{enumerate}
\usepackage{enumitem}
\usepackage{eqparbox}
\usepackage[T1]{fontenc}
\usepackage{flushend}
\usepackage{footnote}
\usepackage{graphicx}
\usepackage{makecell}
\usepackage{mathtools}
\usepackage{mhchem}
\usepackage{mdwmath}
\usepackage{mdwtab}
\usepackage{multirow}
\usepackage{nomencl}
\usepackage{stfloats}
\usepackage{textcase}
\usepackage{threeparttable}
\usepackage{url}
\usepackage{setspace}

\usepackage{etoolbox}
\makeatletter
\patchcmd{\@makecaption}
  {\addvspace{0.5\baselineskip}\egroup}
  {\addvspace{-1\baselineskip}\egroup}
  {}
  {}
\makeatother
\begin{document}

\title{Regional Frequency-Constrained Planning for the Optimal Sizing of Power Systems via Enhanced Input Convex Neural Networks}
\IEEEaftertitletext{\vspace{-1.98\baselineskip}}

\author{Yi Wang,~\IEEEmembership{Member,~IEEE,}
        and Goran Strbac,~\IEEEmembership{Member,~IEEE}
        \thanks{This work was supported by the UK EPSRC project: `Integrated Development of Low-Carbon Energy Systems (IDLES): A Whole-System Paradigm for Creating a National Strategy' (project code: EP/R045518/1) and the Horizon Europe project: `Reliability, Resilience and Defense technology for the griD' (Grant agreement ID: 101075714).}
}
\markboth{IEEE Trans. Sustain. Energy, Accepted for Publication}%
{Shell \MakeLowercase{\textit{et al.}}: Bare Demo of IEEEtran.cls for IEEE Journals}

\maketitle

\begin{abstract}
Large renewable penetration has been witnessed in power systems, resulting in reduced levels of system inertia and increasing requirements for frequency response services. There have been plenty of studies developing frequency-constrained models for power system security. However, most existing literature only considers uniform frequency security, while neglecting frequency spatial differences in different regions. To fill this gap, this paper proposes a novel planning model for the optimal sizing problem of power systems, capturing regional frequency security and inter-area frequency oscillations. Specifically, regional frequency constraints are first extracted via an enhanced input convex neural network (ICNN) and then embedded into the original optimisation for frequency security, where a principled weight initialisation strategy is adopted to deal with the gradient vanishing issues of non-negative weights in traditional ICNNs and enhance its fitting ability. An adaptive genetic algorithm with sparsity calculation and local search is developed to separate the planning model into two stages and effectively solve it iteratively. Case studies have been conducted on three different power systems to verify the effectiveness of the proposed frequency-constrained planning model in ensuring regional system security and obtaining realistic investment decisions.
\end{abstract}
\vspace{-0.28em}
\begin{IEEEkeywords}
Regional frequency security, Optimal sizing, Input convex neural networks, Principled weight initialisation, and Adaptive genetic algorithm.
\end{IEEEkeywords}

\renewcommand{\nomgroup}[1]{%
\ifthenelse{\equal{#1}{A}}{\item[\emph{A.~Abbreviation}]}{%
\ifthenelse{\equal{#1}{B}}{\item[\emph{B.~Indices~and~Sets}]}{%
\ifthenelse{\equal{#1}{C}}{\item[\emph{C.~Parameters}]}{%
\ifthenelse{\equal{#1}{D}}{\item[\emph{D.~Variables}]}{{}}}}}
}
\makenomenclature
\setlength{\nomlabelwidth}{1.95cm} 
\setlength{\nomitemsep}{0.215mm} 
\nomenclature[A01]{RES}{Renewable energy resource}
\nomenclature[A02]{RoCoF}{Rate of change of frequency}
\nomenclature[A04]{ICNN}{Input convex neural network}
\nomenclature[A05]{DNN}{Deep neural networks}
\nomenclature[A06]{PWI}{Principled weight initialisation}
\nomenclature[A07]{AGA}{Adaptive genetic algorithm}
\nomenclature[A08]{UC}{Unit commitment}
\nomenclature[A09]{WT}{Wind turbine}
\nomenclature[A10]{PV}{Photovoltaic}
\nomenclature[A11]{ES}{Energy storage}
\nomenclature[A13]{PFR}{Primary frequency response}
\nomenclature[A14]{EFR}{Enhanced frequency response}
\nomenclature[A15]{COI}{Centre of inertia}
\nomenclature[A16]{OPF}{Optimal power flow}
\nomenclature[B01]{$t \in T$}{Index and set of time steps}
\nomenclature[B02]{$g \in \mathcal{G}$}{Index and set of generators}
\nomenclature[B04]{$k \in \mathcal{ES}$}{Index and set of ESs}
\nomenclature[B05]{$g \in \mathcal{RES}$}{Index and set of RESs}
\nomenclature[B08]{$d \in \mathcal{ED}$}{Index and set of electric demands}
\nomenclature[C01]{$\Delta t$}{Time resolution}
\nomenclature[C02]{$c^{gen}_{g,w}$}{Generation cost of generator $g$ with type $w$}
\nomenclature[C03]{$c^{no}_{g,w}$}{No load cost of generator $g$ with type $w$}
\nomenclature[C04]{$c^{st}_{g,w}$}{Start-up cost of generator $g$ with type $w$}
\nomenclature[C05]{$c^{ic}_{i,t}$}{Electricity cost of interconnector $i$ at time $t$}
\nomenclature[C12]{$P^{ed}_{d,t}$}{Electric load $d$ at time $t$}
\nomenclature[C13]{$\overline{P}_{g}^{w},\underline{P}_{g}^{w}$}{Power output limits of generator $g$ with type $w$}
\nomenclature[C16]{$RU^{w}_{g,k}$}{Ramp-up rate $k$ of generator $g$ with type $w$}
\nomenclature[C17]{$RD^{w}_{g,k}$}{Ramp-down rate $k$ of generator $g$ with type $w$}
\nomenclature[C18]{$\overline{P}_{k}^{es}$}{Power capacity of ES $k$}
\nomenclature[C20]{$\overline{E}_{k}^{es}$}{Energy capacity of ES $k$}
\nomenclature[C21]{$\eta^{es}_{k}$}{Charging/discharging coefficient of ES $k$}
\nomenclature[C22]{$\overline{P}^{res}_{g}$}{Active power capacity of RES $g$}
\nomenclature[C27]{$f_{0}$}{Nominal grid frequency}
\nomenclature[C28]{$\overline{RoCoF}$}{Maximum Rate of Change of Frequency}
\nomenclature[C29]{$\Delta\overline{f}$}{Maximum admissible frequency deviation at the nadir}
\nomenclature[C30]{$\Delta\overline{f}^{ss}$}{Admissible frequency deviation at quasi-steady-state}
\nomenclature[C31]{$\overline{R}^{w}_{g}$}{PFR capacity of generator $g$ with type $w$}
\nomenclature[C32]{$\overline{R}^{es}_{k}$}{EFR capacity of ES $k$}
\nomenclature[C33]{$\overline{V}$}{Maximum permissible voltage}
\nomenclature[C34]{$\underline{V}$}{Minimum permissible voltage}
\nomenclature[C35]{$B_{bp}$}{Susceptance of line $b-p$}
\nomenclature[C36]{$G_{bp}$}{Conductance of line $b-p$}
\nomenclature[C37]{$\overline{S}_{bp}$}{Cpacity limit of line $b-p$}
\nomenclature[C38]{$CI^{w}_{g}$}{Carbon intensity of generator $g$ with type $w$}

\nomenclature[D01]{$P^{w}_{g,t},Q^{w}_{g,t}$}{Active/Reactive power output of generator $g$ with type $w$}
\nomenclature[D03]{$y^{w}_{g,t}$}{Commitment state of generator $g$ with type $w$}
\nomenclature[D04]{$y^{w,sg}_{g,t}$}{`Generating' state of generator $g$ with type $w$}
\nomenclature[D05]{$y^{w,st}_{g,t}$}{Startup variable for generator $g$ with type $w$}
\nomenclature[D06]{$y^{w,sd}_{g,t}$}{Shutdown variable for generator $g$ with type $w$}
\nomenclature[D09]{$P_{k,t}^{c}$}{Charging power of ES $k$}
\nomenclature[D10]{$P_{k,t}^{d}$}{Discharging power of ES $k$}
\nomenclature[D12]{$E_{k,t}^{es}$}{Energy content of ES $k$}
\nomenclature[D13]{$u^{es}_{k,t}$}{Binary indicating the charging/discharging status of ES $k$ at time $t$}
\nomenclature[D14]{$P^{res}_{g,t}$}{Active power output of RES $g$ at time $t$}
\nomenclature[D15]{$P^{ic}_{i,t}$}{Active power of interconnector $i$ at time $t$}
\nomenclature[D18]{$H_{t}$}{System inertia at time $t$}
\nomenclature[D19]{$P^{GL}_{t}$}{Largest power infeed loss at time $t$}
\nomenclature[D20]{$R^{w}_{g,t}$}{PFR of generator $g$ with type $w$ at time $t$}
\nomenclature[D21]{$R^{es}_{k,t}$}{EFR of ES $k$ at time $t$}
\nomenclature[D22]{$V_{b,t}$}{Voltage of bus $b$ at time $t$}
\nomenclature[D23]{$\delta_{bp,t}$}{Voltage angle difference between bus $b$ and bus $p$ at time $t$} 
\nomenclature[D24]{$P^{ex}_{b,t}$}{Active power exchange between bus $b$ and other buses at time $t$} 
\nomenclature[D25]{$Q^{ex}_{b,t}$}{Reactive power exchange between bus $b$ and other buses at time $t$}
\nomenclature[D26]{$P_{bp,t}$}{Active power of line $b-p$ at time $t$}
\nomenclature[D27]{$Q_{bp,t}$}{Reactive power of line $b-p$ at time $t$}
\printnomenclature

\section{Introduction}
\label{sec:I}
\subsection{Research Background and Motivation}
\label{sec:I.A}
\IEEEPARstart{I}n response to the low-carbon requirement, power systems are transitioning significantly from synchronous fossil fuel resources to converter-interfaced renewable energy resources (RESs) such as wind and solar \cite{strbac2019cost}. However, the high penetration of RESs can cause the reduced inertia level of future power systems, which increases the risk of frequency instability issues \cite{teng2015stochastic}. To ensure frequency security, system operators have to procure various frequency response services from energy-intensive industries, leading to much research on frequency-constrained power system operation and planning problems, where frequency nadir and Rate of Change of Frequency (RoCoF) are the keys to ensuring frequency security.

Regarding frequency-constrained operations, intensive studies have focused on deducing mathematical constraints from swing equations for frequency security and embedding them into the original optimisation problem \cite{zhang2020modeling,badesa2019simultaneous,yin2021frequency}. In general, the above research considers a single-bus representation assuming that all the generators move coherently as a single lumped mass. However, recent studies have shown that the Centre Of Inertia (COI) representation can be inaccurate in modern power systems, where the unbalanced allocation of RESs at different areas creates a non-uniform distribution of inertia, leading to spatial gradients and distinct regional frequencies \cite{part1}. In this context, even though the frequency requirement under COI is satisfactory, the RoCoF or nadir of some regions may still go beyond the limits, causing frequency collapse.

\subsection{Literature Review}
\label{sec:I.B}
To guarantee regional frequency stability, an analytical-numerical approach based on multi-regions is proposed in \cite{badesa2021conditions}, capturing the combining evolution of COI and inter-area oscillations. However, this approach over-simplifies the problem, neglecting the impact of disturbance propagation on regional inertial response. In this context, a location-based RoCoF-constrained model is proposed in \cite{tuo2022security} to capture regional frequency characteristics, counteract the impact of system oscillations, and guarantee RoCoF security. In \cite{rabbanifar2020frequency}, an area RoCoF expression is presented to estimate the regional frequency responses following a generator loss. In \cite{liu2023rocof}, the analytical expressions of the nodal RoCoF after the disturbance are derived, considering location and disturbance severity. However, the above papers \cite{tuo2022security,rabbanifar2020frequency,liu2023rocof} mainly focus on RoCoF-based frequency dynamics while ignoring the spatial difference of frequency nadir. Going further, the above linearisation methods \cite{badesa2021conditions,tuo2022security,rabbanifar2020frequency,liu2023rocof} together with \cite{zhang2020modeling,badesa2019simultaneous,yin2021frequency} for frequency security can only derive approximate constraints based on certain operation states or assumptions. However, high RES penetration leads to a substantial diversity in operation states, making it challenging for the generality of these methods. Furthermore, a fundamental challenge is that the above mathematical formulations highly rely on the precise system and topology parameter estimations, while power systems normally suffer from a lack of observability and system parameters may be unknown or inaccurate \cite{chen2020data}. This limits the applicability of analytical approaches in real-world scenarios.

In recent years, system operators have had access to a large amount of operation data due to the trend of digitalisation. As the power system operating status (e.g., regional frequency security) is greatly associated with operating points, system conditions can be approximated using data-driven approaches instead of analytic formulations \cite{10104114}. In this context, due to the powerful capability of feature extraction from large datasets, data-driven and deep learning techniques offer significant advantages by addressing the above challenges. On one hand, unlike traditional model-based approaches that require exact knowledge of system parameters, data-driven models based on deep neural networks (DNNs) can learn directly from historical or real-time data, making them more robust to estimation errors in system parameters. On the other hand, deep learning methods can adapt to varying operational conditions and handle the diversity introduced by different operation states, enabling more flexible and generalisable solutions as well as accelerating the solution process.
Therefore, by leveraging large-scale datasets, DNNs can be regarded as a scalable and efficient way to capture complex relationships between operational states and system dynamics (e.g., frequency security). In \cite{10104114}, conservative sparse neural networks are adopted to learn frequency dynamics and then embedded into frequency-constrained unit commitment (UC) problems for reliable system operation. In \cite{xia2024efficient}, an iterative method is proposed to embed DNN-based stability constraints into operation problems solved in the real-time stage. In \cite{zhang2021encoding}, a data-driven framework is developed for frequency-constrained UC, where DNNs are trained to predict the frequency response using high-fidelity simulation data. Therefore, it could be interesting to extract regional frequency constraints via DNNs and then embed them into the original optimisation for frequency security; nevertheless, this has not yet been appropriately investigated in the existing literature. 

Furthermore, to be solved efficiently, the embedded DNN should be computationally tractable (e.g., convex) or can return effective cutting plane information \cite{wu2023transient}. However, DNNs used in \cite{10104114,xia2024efficient,zhang2021encoding} fail to meet this requirement due to their non-linear nature. Compared with DNNs, input convex neural networks (ICNNs) have been regarded as a promising solution to output convex constraints for linear programming, due to their specific settings on the weights and skip-connections. There have been papers focused on training ICNNs and then embedding them into voltage regulation and optimal power flow (OPF) problems. For instance, in \cite{chen2020data}, ICNNs are used for distribution network modelling and voltage regulation problems within a convex optimisation framework, enabling a tractable formulation to find reactive power injections. In \cite{rosemberg2024learning}, ICNN architectures are trained to approximate the value function of large-scale OPF problems, while \cite{zhang2021convex} combines ICNNs with Karush–Kuhn–Tucker optimality conditions to solve DC OPF problems with guaranteed generalisation. However, the above works primarily focus on the static power system operation problems without considering post-fault dynamics. To address this, ICNNs are trained in \cite{wu2023transient} to learn the transient function that maps pre-fault operation conditions to the transient stability. Despite these advancements, the fitting ability and convergence capability of traditional ICNNs are normally compromised, leading to conservative solutions, due to the hard limits on non-negative weights and non-decreasing activation functions \cite{hoedt2024principled}. Note that abstracting frequency-related constraints needs a massive dataset to characterise the distribution of operation points or system conditions, necessitating a strong fitting ability of ICNNs.

From the perspective of power system planning, most existing literature only focuses on the economic perspective, i.e., minimisation of investment costs. There are several papers considering frequency security in investment decisions. For instance, in \cite{nakiganda2022stochastic}, an investment planning model is developed for the design of a resilient microgrid, capturing dynamic frequency response during the islanding period. However, the UC of generators is not considered. In \cite{zhao2023scalable}, a stochastic planning model capturing frequency dynamics is proposed for the optimal sizing of batteries in power systems; nevertheless, the investment decisions of generators are not considered. In \cite{li2021frequency}, a frequency-constrained planning method is proposed for a power system with high RES penetration, capturing the frequency response support from wind farms. However, this paper employs a simplified UC model based on aggregated generation, which may suffer a flexibility inadequacy problem and make it difficult to accurately estimate the number of online generators, leading to inaccurate inertia calculation. Finally, the above papers \cite{nakiganda2022stochastic,zhao2023scalable,li2021frequency} only consider the planning problem under uniform frequency concept, while regional frequency security has been ignored, which can be impractical. Note that investment decisions based on uniform frequency security may not be able to ensure regional frequency security.
\vspace{-1.38em}

\subsection{Paper Contributions}
\label{sec:I.C}
To fill the above research gaps, this paper aims to propose a novel ICNN-based planning model for the optimal sizing of power systems capturing regional frequency security. The detailed contributions are summarised below:

1) An enhanced ICNN is designed and trained to map pre-fault operation conditions to regional frequency security. The trained ICNN is encoded as a linear problem based on its convexity and embedded into the optimisation model. 

2) A novel principled initialisation strategy is introduced to address gradient vanishing issues in conventional ICNNs with non-negative weights, eliminating the need for skip connections and reducing model complexity.

3) A planning framework is proposed for optimal sizing of power systems, capturing regional frequency security requirements of multi-region systems. A detailed UC model considering different ramping rates is developed to account for the scheduling behaviours of each individual generator.

4) To reduce model complexity, an adaptive genetic algorithm (AGA) with sparsity calculation and local search is proposed to separate the planning model into two levels for efficient solving. Post-fault dynamic simulations are integrated to verify the frequency security of obtained solutions.
\vspace{-0.38em}

\subsection{Paper Organisation}
\label{sec:I.D}
The rest of this paper is organised as follows. Section \ref{sec:II} provides the general mathematical formulations of the proposed regional frequency-constrained optimisation model. Section \ref{sec:III} introduces the enhanced ICNN model based on principled weight initialisation, while Section \ref{sec:IV} presents the final mathematical formulation of the planning model and related solving procedure based on AGA. In Section \ref{sec:V}, case studies are carried out and analysed on three power systems, including the IEEE 6-bus system, the 14-bus GB power system, and the modified IEEE 118-bus system. Section \ref{sec:VI} draws the conclusions of this paper.

\begin{figure}[t!] 
\centering  
\includegraphics[width=0.485\textwidth]{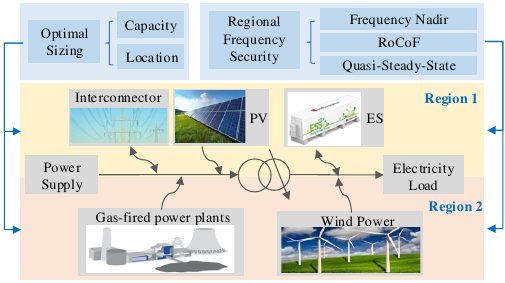}
\vspace{-1.08em}
\caption{Illustration of the proposed frequency-constrained planning model capturing regional inter-area frequency oscillations.}
\label{fig:problem}
\vspace{-1.28em}
\end{figure}

\section{General Mathematical Formulation of the Proposed Planning Model}
\label{sec:II}
\subsection{Problem Descriptions}
\label{sec:II.A}
This paper focuses on the optimal sizing problem of power systems with the consideration of post-fault regional frequency security. As depicted in Fig. \ref{fig:problem}, the sizing and positioning of different types of conventional generators, wind turbines (WTs), photovoltaics (PVs), and energy storage systems (ESs) are considered at the planning stage. To ensure secure system operations of multi-regions, post-fault transient system dynamics involving the combining evolution of COI and inter-area frequency oscillations are captured in investment decisions.

Specifically, the key distinction between uniform frequency models and the proposed regional frequency model lies in how they address inter-area oscillations and regional frequency variations. Traditional uniform frequency models typically assume that the entire power system operates at a single and coherent frequency, simplifying frequency stability dynamics by treating the grid as a unified entity. In contrast, the regional frequency model recognises that different regions in power systems may exhibit distinct frequency dynamics, especially in systems with high penetration of non-synchronous RESs. Therefore, it can account for spatial variations in system inertia and inter-area frequency oscillations, providing a more accurate representation of frequency stability in power systems with multiple regions. By modelling each region, the regional frequency model captures regional frequency deviations and allows for more realistic co-optimisation of ancillary services, such as inertia, primary frequency response (PFR), enhanced frequency response (EFR), etc.

\subsection{General Mathematical Formulations}
\label{sec:II.B}
In summary, the planning model should consider various operation constraints, aiming to minimise the system cost while meeting regional frequency security requirements. Three sets of constraints should be considered, including the UC constraints of conventional generators, power flow constraints, and frequency-related constraints. Detailed mathematical formulations can be found below:

\subsubsection{Generator Model}
\label{sec:II.B.1}
\begin{figure}[h!]
\centering
{\includegraphics[width=0.46\textwidth]{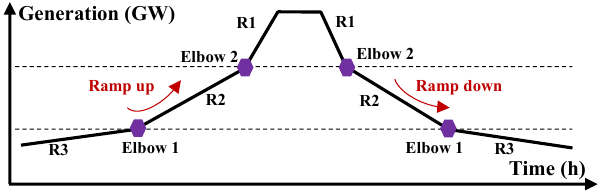}}
\vspace{-0.38em}
\caption{Illustration of one generator with three different ramping rates.}
\label{fig:ramp}
\end{figure}

The generation and ramp-up/down limits of generators are modelled as
\begin{equation}\label{eq:gen energy range}
    y^{w}_{g,t}\underline{P}_{g}^{w} \leq P_{g,t}^{w} \leq y^{w}_{g,t}\overline{P}_{g}^{w}, \forall w,g,t,
\end{equation}
\begin{equation}\label{eq:gen ramp up range}
\begin{split}
     \!P_{g,t}^{w} \!- P_{g,t-1}^{w} \leq RU^{w}_{g,k}y^{w}_{g,t}, if~P_{g,t-1}^{w} \!\in\! [E^{w}_{g,k},E^{w}_{g,k+1}), \\ \forall w,g,k,t,
\end{split}
\end{equation}
\begin{equation}\label{eq:gen ramp down range}
\begin{split}
     \!\!P_{g,t-1}^{w} \!- P_{g,t}^{w} \leq RD^{w}_{g,k}y^{w}_{g,t}, if~P_{g,t-1}^{w} \in [E^{w}_{g,k},E^{w}_{g,k+1}), \\ \forall w,g,k,t,
\end{split}
\end{equation}
\begin{equation}\label{eq:gen on}
y^{w}_{g,t} = y^{w}_{g,t-1} + y^{w,sg}_{g,t-1}-y^{w,sd}_{g,t},\forall w,g,t,
\end{equation}
\begin{equation}\label{eq:gen start}
y^{w,sg}_{g,t} = y^{w,st}_{g,t-T^{st}_{g}}, \forall w,g,t,
\end{equation}
\begin{equation}\label{eq:gen mdt}
y^{w,st}_{g,t} \leq (1-y^{w}_{g,t-1})-\!\!\!\sum_{\tau=t-T^{d}_{g}}^{t}y^{w,sd}_{g,\tau},\forall w,g,t,
\end{equation}
\begin{equation}\label{eq:gen mut}
y^{w,sd}_{g,t} \leq y^{w}_{g,t-1}-\!\!\!\!\sum_{\tau=t-T^{u}_{g}}^{t}y^{w,sg}_{g,\tau},\forall w,g,t,
\end{equation}
where \eqref{eq:gen energy range} corresponds to the output limits of generator $g$ with type $w$. Constraints \eqref{eq:gen ramp up range}-\eqref{eq:gen ramp down range} refer to the $k^{th}$ ramp-up and ramp-down limits $RU^{w}_{g,k},RD^{w}_{g,k}$ of generator $g$ with type $w$ at a specific generation range $[E^{w}_{g,k},E^{w}_{g,k+1})$, respectively. In other words, as depicted in Fig. \ref{fig:ramp}, the ramp-up/down rates can change for each time step $t$, depending on the current generation level \cite{gridcode}. It is worth noting that conventional thermal generators may have different ramping rates based on how long they have been running (e.g., cold, warm, or hot start conditions). For instance, when thermal generators are connected to the grid over a certain time period (e.g., moving from low to higher output levels), they typically operate in more stable and efficient zones. Therefore, the mechanical and thermal systems can handle faster adjustments without exceeding safe operational limits, which may allow for an increase in ramping rates. Constraint \eqref{eq:gen on} describes that the commitment state of generator $g$ at time $t$ is `on' if it was generating at time $t-1$ or has started generating, unless it has been shut down at time $t$. Constraint \eqref{eq:gen start} shows that generator $g$ starts generating after it started up for `$T^{st}_{g}$' periods. The conditions of start-up and shut-down of generator $g$ are expressed in \eqref{eq:gen mdt}-\eqref{eq:gen mut}, respectively.

In addition, ESs and RESs are modelled as
\begin{equation}\label{eq:ev model charge}
    0 \leq P^{c}_{k,t} \leq u^{es}_{k,t} \overline{P}^{es}_{k}, \forall k,t,
\end{equation}
\begin{equation}\label{eq:ev model discharge}
    (u^{es}_{k,t}-1) \overline{P}^{es}_{k} \leq P^{d}_{k,t} \leq 0, \forall k,t,
\end{equation}
\begin{equation}\label{eq:ev storage energy}
    \underline{E}^{es}_{k} \leq E_{k,t}^{es} \leq \overline{E}^{es}_{k}, \forall k,t,
\end{equation}
\begin{equation}\label{eq:ev storage soc}
\begin{split}
\!\!\!\!\!\!\!E_{k,t+1}^{es} \!=\! E_{k,t}^{es} + (P^{c}_{k,t}\eta_{k}^{es} \!+\! P^{d}_{k,t}/\eta_{k}^{es}) \Delta t, \forall k,t,
\end{split}
\end{equation}
\begin{equation}\label{acpowerres}
    0 \leq P^{res}_{g,t} \leq \alpha_{g,t}\overline{P}^{res}_{g}, \forall g,t,
\end{equation}
where the power charging/discharging and energy content of ES $k$ are limited in \eqref{eq:ev model charge}-\eqref{eq:ev storage soc}. $u^{es}_{k,t}$ indicates the status of ES $k$ at time $t$ ($u^{es}_{k,t}=1$ if charging; 0, otherwise). Constraint \eqref{acpowerres} limits the active power of RES $g$ at time $t$, which is affected by weather conditions (e.g., wind speed) contained in $\alpha_{g,t}$. 

\subsubsection{Network Operation}
\label{sec:II.B.3}
The power system is modelled by a linearised AC OPF \cite{wang2021three}, expressed as
\begin{equation}\label{eq:acopf_bal1}
\begin{split}
\sum_{i \in B_{ic}}\!\!P^{ic}_{i,t}+\sum_{g \in B_{eg}}\sum_{w \in \mathcal{W}}\!\!P_{g,t}^{w} + \!\!\sum_{g \in B_{res}}\!\!P_{g,t}^{res} = P^{ex}_{b,t} + \\ \sum_{d \in B_{ed}}\!P^{ed}_{d,t} + \sum_{k \in B_{es}}(P^{c}_{k,t} - P^{d}_{k,t}), \forall b,t,
\end{split}
\end{equation}
\begin{equation}\label{eq:acopf_bal2}
    \sum_{g \in B_{eg}}\sum_{w \in \mathcal{W}}Q_{g,t}^{w} = Q^{ex}_{b,t} + \sum_{d \in B_{ed}} Q^{ed}_{d,t}, \forall b,t,
\end{equation}
\begin{equation}\label{eq:acopf_classical1}
    P^{ex}_{b,t} = \sum_{bp \in \mathcal{L}}P_{bp,t} + (\sum_{p \in \mathcal{B}} G_{bp}) V^2_{b,t}, \forall b,bp,t,
\end{equation}
\begin{equation}\label{eq:acopf_classical2}
    Q^{ex}_{b,t} = \sum_{bp \in \mathcal{L}}Q_{bp,t} - (\sum_{p \in \mathcal{B}}B_{bp}) V^2_{b,t}, \forall b,bp,t,
\end{equation}
\begin{equation}\label{eq:acopf_classical3}
\begin{split}
    P_{bp,t} = G_{bp} (V^2_{b,t}-V^2_{p,t})/2 - B_{bp} \delta_{bp,t} + P^{lo}_{bp,t}, \forall bp,t,
\end{split}    
\end{equation}
\begin{equation}\label{eq:acopf_classical4}
\begin{split}
    \!\!Q_{bp,t}\!=\! -B_{bp} (V^2_{b,t}-V^2_{p,t})/2 - G_{bp} \delta_{bp,t} + Q^{lo}_{bp,t},\forall bp, t,
\end{split}
\end{equation}
\begin{equation}\label{eq:acopf_vol}
    \underline{V}^2 \leq V^2_{b,t} \leq \overline{V}^2, \forall b,t,
\end{equation}
\begin{equation}\label{eq:acopf_thermal}
    P_{bp,t}^2 + Q_{bp,t}^2 \leq \overline{S}^2_{bp}, \forall bp, t,
\end{equation}
where constraints \eqref{eq:acopf_bal1}-\eqref{eq:acopf_bal2} indicate the active and reactive power balances at bus $b$. Sets $B_{ic}$, $B_{eg}$, $B_{ed}$, $B_{res}$, and $B_{es}$ correspond to the interconnector (IC) $i$, generator $g$, load $d$, RES $g$, and ES $k$ connected with bus $b$, respectively. Nodal power exchanges $P^{ex}_{b,t},Q^{ex}_{b,t}$ at bus $b$ can be linearised by \eqref{eq:acopf_classical1}-\eqref{eq:acopf_classical4}, where $P^{lo}_{bp,t},Q^{lo}_{bp,t}$ are active and reactive power losses that are linearised by loss factors \cite{yang2017linearized}. The nodal voltage and line capacity limits are expressed in \eqref{eq:acopf_vol} and \eqref{eq:acopf_thermal}, respectively. Furthermore, the overall carbon emissions are limited by 
\begin{equation} \label{eq:carbon}
    \sum_{w \in \mathcal{W}}\sum_{g \in \mathcal{G}}\sum_{t \in T} P^{w}_{g,t} CI^{w}_{g} \leq \overline{CO_{2}}\sum_{d \in \mathcal{ED}}\sum_{t \in T} P^{ed}_{d,t},
\end{equation}
where $CI^{w}_{g}$ corresponds to the carbon intensity factor of generator $g$ with type $w$. $\overline{CO_{2}}$ is the carbon target, ensuring that the overall carbon emissions do not exceed the regulated limit of carbon emissions.
\vspace{-0.28em}

\subsubsection{Regional Frequency-related Constraints}
\label{sec:II.B.4}
The RoCoF and frequency nadir for region $n \in \mathcal{N}$ as well as quasi-steady-state constraints can be expressed as
\begin{equation}\label{eq:tso rocof}
    |\frac{d\Delta f_{n}(t)}{dt}| \leq \overline{RoCoF}, \forall n,t,
\end{equation}
\begin{equation}\label{eq:tso nadir}
   |\Delta f_{n}(t)| \leq \Delta\overline{f}, \forall n,t,
\end{equation}
\begin{equation}\label{eq:tso qss}
    \sum_{g \in \mathcal{G}}R^{w}_{g,t}+\sum_{k \in \mathcal{ES}}R^{es}_{k,t} \geq P_{t}^{GL}-\Delta \overline{f}^{ss}\sum_{n \in \mathcal{N}}D_{n}P^{ed}_{n}, \forall t,
\end{equation}
where $\overline{RoCoF}$ are the maximum admissible RoCoF \cite{badesa2019simultaneous}. $\Delta \overline{f}$ is the maximum admissible frequency deviation, while $\Delta \overline{f}^{ss}$ is admissible frequency deviation at quasi-steady-state. $\Delta f_{n}(t)$ denotes the frequency deviations in region $n$. $D_n$ and $P^{ed}_{n}$ represent the load damping factor and demand level in region $n$, respectively. Note that constraint \eqref{eq:tso qss} ensures that the sum of frequency response is greater than the largest loss so that frequency drop can be finally arrested. {\color{black}Similarly, the unpredictability associated with the substantial integration of RESs can also be managed by including related operating reserve constraints, such as:
\begin{equation}\label{eq:reserve_new}
    \sum_{g \in \mathcal{G}}OR^{w}_{g,t}+\sum_{k \in \mathcal{ES}}OR^{es}_{k,t} \geq \epsilon RES_{t}, \forall t,
\end{equation}
where $OR^{w}_{g,t}$ and $OR^{es}_{k,t}$ correspond to the operating reserve from generators and ESs, respectively. $\epsilon$ is the forecasting error of RES, following the same practice in \cite{zhang2018whole}.}

Schedules of frequency services are modelled as
\begin{equation}\label{eq:tso inertia}
    H_{t}=\sum_{g \in \mathcal{G}}\sum_{g \in \mathcal{W}}H^{w}_{g}\overline{P}^{w}_{g}y^{w}_{g,t}-H^{GL}P_{t}^{GL}, \forall t,
\end{equation}
\vspace{-0.48em}
\begin{equation}\label{eq:ic gen loss}
 P^{GL}_{t} \geq P^{w}_{g,t}, \forall w,g,t,
\end{equation}
\vspace{-0.78em}
\begin{equation}\label{eq:fr_dg_limit}
    0 \leq R^{w}_{g,t} \leq y^{w}_{g,t}\overline{P}^{w}_{g}-P^{w}_{g,t}, \forall w,g,t,
\end{equation}
\begin{equation}\label{eq:fr_dg_max}
    0 \leq R^{w}_{g,t} \leq \overline{R}^{w}_{g}, \forall w,g,t,
\end{equation}
\begin{equation}\label{eq:ev efr_limit}
    \!\!\!0 \leq R^{es}_{k,t} \leq (1 - u^{es}_{k,t})\overline{P}^{es}_{k}+P^{d}_{k,t}+P^{c}_{k,t}, \forall k,t,
\end{equation}
\begin{equation}\label{eq:ev efr_limit_2}
     0 \leq R^{es}_{k,t} \leq \overline{R}^{es}_{k}, \forall k,t,
\end{equation}
where inertia $H_{t}$ is calculated in \eqref{eq:tso inertia} by aggregating the inertia of generators, except for the largest power infeed loss $P_{t}^{GL}$. Note that the largest power infeed loss refers to the maximum potential loss of generation capacity in a power system resulting from the sudden failure of the largest generator, IC, or other critical source of power \cite{badesa2019simultaneous,zhang2020modeling}, which can be expressed in \eqref{eq:ic gen loss}. The PFR and EFR quantity of generator $g$ and ES $k$ are limited by \eqref{eq:fr_dg_limit}-\eqref{eq:fr_dg_max} and \eqref{eq:ev efr_limit}-\eqref{eq:ev efr_limit_2}, respectively. Similarly, constraint set \eqref{eq:tso inertia}-\eqref{eq:ev efr_limit_2} can be duplicated to account for the largest loss of demand $P^{LL}_{t}$. For instance, the power exporting behaviours of ICs may cause a loss of demand.

\subsubsection{Objective Function}
\label{sec:II.B.4}
The objective includes annuitized investment costs and operation costs, which are expressed as
\begin{equation} \label{eq:obj}
\begin{split}
\min &\sum_{w \in \mathcal{W}} G^{cap}_{w} (C^{gen}_{w}AF^{gen}_{w}+C^{fix,gen}_{w})\\
&+\sum_{g \in \mathcal{RES}} RES^{cap}_{g} (C^{res}_{g}AF^{res}_{g}+C^{fix,res}_{g})\\
&+\sum_{k \in \mathcal{ES}} ES^{cap}_{k} (C^{es}_{k}AF^{es}_{k}+C^{fix,es}_{k})+F^{op},
\end{split}
\end{equation}
where the investment costs of generators, RESs (WTs and PVs), and ESs consider annuity factors (e.g., $AF^{gen}_{w}$) and fixed operating and maintenance cost (e.g., $C^{fix,gen}_{w}$) \cite{zhang2018whole}. The annual operation cost $F^{op}$ is defined as the operating cost of generators and ICs, which can be written as
\begin{equation}\label{eq:tso obj}
\begin{split}
    F^{op} &= \sum_{w \in \mathcal{W}}\sum_{g \in \mathcal{G}}\sum_{t \in T} c^{gen}_{g,w} P_{g,t}^{w}+\sum_{i \in \mathcal{IC}}\sum_{t \in T}c^{ic}_{i,t}P^{ic}_{i,t}\\
    &+\sum_{w \in \mathcal{W}}\sum_{g \in \mathcal{G}}\sum_{t \in T} (c^{no}_{g,w} y^{w}_{g,t} + c^{st}_{g,w} y^{w,st}_{g,t}),
\end{split}
\end{equation}
where $c^{gen}_{g,w}$, $c^{st}_{g,w}$, and $c^{no}_{g,w}$ correspond to the marginal generation cost, start-up cost, and no load cost of generator $g$ with type $w$, respectively. $c^{ic}_{i,t}$ is the electricity price of IC $i$.

\subsection{Problem Challenges}
\label{sec:II.C}
To maintain post-fault system security, frequency-related constraints \eqref{eq:tso rocof}-\eqref{eq:tso qss} should be included in the planning model. However, directly solving the above planning model faces the two following challenges:
\begin{itemize}
    \item Regional frequency constraints \eqref{eq:tso rocof}-\eqref{eq:tso qss} involve post-fault system dynamics expressed as differential equations and inter-area frequency oscillations, which are highly non-linear and cannot be directly included as constraints. The consideration of both under and over frequency security requirements further increases the model complexity.
    \item To account for accurate inertia calculation and mimic real-world operation practices, the proposed planning model considers a detailed operation problem including the schedules of each generator (e.g., \eqref{eq:gen energy range}-\eqref{eq:gen mut}). Specifically, different ramp-up/down rates are applied to each generator, significantly increasing the model complexity.
\end{itemize}

\section{Generating Regional Frequency Constraints}
\label{sec:III}
This section addresses the first challenge mentioned above by introducing enhanced ICNNs with a novel principled weight initialisation (PWI) strategy to formulate regional frequency security requirements \eqref{eq:tso rocof}-\eqref{eq:tso qss} as convex constraints.
\vspace{-1.38em}

\subsection{Preliminaries of Regional Frequency Security}
\label{sec:IV.A}
Let's first understand the dynamics of post-fault frequency evolution in multi-region systems. In general, the concept of `regions' is defined from the electromechanical coupling in a system to consider intra-area oscillations. In detail, the power system is modelled as a set of regions that do not swing coherently but are connected through synchronous AC transmission lines. The post-fault frequency in a multi-region system can be expressed as the following coupled swing equations considering geographical gradients in frequency and inter-region power transfer:
\begin{equation}\label{eq:coupled swing}
\begin{cases}
    2H_{1} \cdot \frac{d\Delta f_{1}(t)}{dt}~+\!\!\!\!\!&D_{1}\cdot P^{ed}_{1} \cdot \Delta f_{1}(t) = PFR_{1}(t)\\&+EFR_{1}(t)-P_{1}^{GL} + \Delta P_{1}^{im}(t),\\
    2H_{2} \cdot \frac{d\Delta f_{2}(t)}{dt}~+\!\!\!\!\!&D_{2}\cdot P^{ed}_{2} \cdot \Delta f_{2}(t) = PFR_{2}(t)\\&+EFR_{2}(t)-P_{2}^{GL} + \Delta P_{2}^{im}(t),\\
    \cdot \cdot \cdot \\
    2H_{n} \cdot \frac{d\Delta f_{n}(t)}{dt}~+\!\!\!\!\!&D_{n}\cdot P^{ed}_{n} \cdot \Delta f_{n}(t) = PFR_{n}(t)\\&+EFR_{n}(t)-P_{n}^{GL} + \Delta P_{n}^{im}(t),
\end{cases}
\end{equation}
where $\Delta P_{n}^{im}(t)$ represents the difference between the power imported after outage and before outage, demonstrating the deviation from the steady state power being imported to region $n$ after one outage so that a frequency equilibrium can be restored in the whole network \cite{badesa2021conditions}. $H_{n}$ denotes the inertia level in region $n$, while $P_{n}^{GL}$ indicates the potential power infeed loss in region $n$. $PFR_{n}$ and $EFR_{n}$ represent the procured PFR and EFR services in region $n$, respectively. It is worth noting that the above coupled swing equations \eqref{eq:coupled swing} are the extension of the single-bus swing equation to consider distinct regions, which can be reduced to the single-bus swing equation if the impedance of the transmission lines tends to zero \cite{part1}. In this context, the key criterion for region partition is the impedance of the transmission line, e.g., the power system can be partitioned into two regions when the impedance between these two regions is large enough. In detail, higher line impedance weakens the coupling strength between regions, allowing each region to respond to disturbances more independently. This results in significant inter-area frequency oscillations, as each region can exhibit distinct oscillation phases and possibly different amplitudes (non-coherent swinging) rather than unified and coherent swing.

The ramping behaviours of PFR and EFR in region $n$ can be defined as \cite{badesa2019simultaneous}:
\begin{equation}\label{eq:pfr}
PFR_{n}(t)=
\begin{cases}
\sum_{g \in \mathcal{G}}R^{w}_{g,n} \cdot t/T^{gen},~\text{if}~t \leq T^{gen},\\ 
\sum_{g \in \mathcal{G}}R^{w}_{g,n},~\text{if}~t > T^{gen},\\ 
\end{cases}
\end{equation}
\begin{equation}\label{eq:efr}
EFR_{n}(t)=
\begin{cases}
\sum_{k \in \mathcal{ES}}R^{es}_{k,n} \cdot t/T^{es}, ~\text{if}~t \leq T^{es},\\ 
\sum_{k \in \mathcal{ES}}R^{es}_{k,n},~\text{if}~t > T^{es}.\\ 
\end{cases}
\end{equation}
where $T^{es},T^{gen}$ correspond to the FR delivery speed from ESs and thermal generators, respectively. Specifically, the post-

\!\!\!\!\!\!fault power imported from any region $n$ can be expressed as
\begin{equation}\label{eq:post import}
\!\! \Delta P_{n}^{im}(t) \!=\! \sum_{m \in \mathcal{N^r}}T_{n,m}[\int_{0}^{t}\Delta f_{n}(\tau)d\tau\!-\!\int_{0}^{t}\Delta f_{m}(\tau)d\tau],
\end{equation}
where $\mathcal{N^r}$ is the set of neighbour regions of region $n$. Note that a negative value of $\Delta P_{n}^{im}(t)$ means that region $n$ is exporting power to other regions. The electrical stiffness $T_{n,m}$ of the transmission line \cite{badesa2021conditions} can be defined as
\begin{equation}\label{eq:stiffness}
T_{n,m}=2\pi \cdot \frac{V_{n}V_{m}}{X_{n,m}} \cdot \cos(\delta_{n}-\delta_{m}),
\end{equation}
where $X_{n,m}$ and $\delta_{n}$ are the reactance and voltage phase angle.

As aforementioned, regional frequency security includes the analysis of inter-area frequency dynamics, which are highly non-linear and cannot be directly included in optimisation. Even though there have been papers (e.g., \cite{badesa2021conditions}) deducing mathematical constraints for regional frequency deviation analysis, traditional linearisation methods can only derive approximate constraints based on certain operation states. However, high RES penetration leads to a substantial diversity in operation states, making it challenging to apply these methods to power system operation and planning problems. Additionally, the conditional constraints developed in \cite{badesa2021conditions} for regional frequency nadir involve large amounts of integer variables, significantly increasing model complexity, causing heavy computational burden, and easily leading to conservative solutions.

\subsection{ICNN-based Regional Frequency Security Constraints}
\begin{figure}[t!]
\centering
{\includegraphics[width=0.44\textwidth]{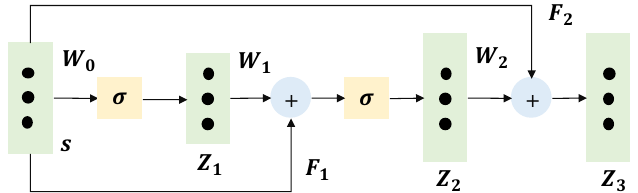}}
\vspace{-0.38em}
\caption{Structure of a two-layer input convex neural network.}
\vspace{-0.68em}
\label{fig:icnn}
\end{figure}

In this section, ICNNs are introduced to extract regional frequency constraints and then embed them into the original optimisation. As shown in Fig. \ref{fig:icnn}, ICNNs consist of an input layer, several hidden layers, and an output layer. $s$, $W$ and $F$ denote the input vector and non-negative weight matrices of ICNNs, respectively. $\phi(\cdot)$ refers to the activation functions. Unlike DNNs, ICNNs link the input layer to each hidden layer as well as the output layer via weight matrix $F$, which are realised as skip-connections for representational power.

\subsubsection{Encoding of Trained ICNN}
To make sure that the output of ICNNs (e.g., $z_{3}$ in Fig. \ref{fig:icnn}) is a convex function over input $s$, two conditions must be satisfied: 1) weights $W_i, \forall i=1:L-1$ are non-negative and 2) all activation functions are convex and non-decreasing \cite{amos2017input}. The first condition is normally achieved using weight clipping or projected gradient algorithms, while \texttt{ReLU} can be an excellent alternative for $\phi(\cdot)$ to satisfy the second condition. The detailed proof of ICNN convexity given the two propositions can be found in \cite{amos2017input}.

To train ICNNs for regional frequency security, data (e.g., inertia, PFR, EFR, damping, demand, largest loss, region, etc.) are fed into the input layer and then conveyed through hidden layers to the output layer. Subsequently, training losses are calculated based on specific loss functions and then passed backwards. This process repeats until it converges. Afterwards, the well-trained ICNN can establish the underlying relation between input and output. Specifically, based on the connection shown in Fig. \ref{fig:icnn} and the adopted activation function (e.g., \texttt{ReLU}), the relation between the input and output of each layer can be expressed as follows:
\begin{equation}\label{eq:icnn structure}
\!\!\!\begin{cases}
z_1=\phi(s^T W_{0} + b_{0})=\max(s^T W_{0} +b_{0},0),\\
z_{i+1}=\phi(z_{i}^{T} W_{i} + s^{T} F_{i} + b_{i})\\
=\max(z_{i}^{T} W_{i} + s^{T} F_{i} + b_{i} +b_{0},0), \forall i \in [1,L\!-\!2],\\
z_{L}=\phi(z_{L-1}^{T} W_{L-1} + s^{T} F_{L-1} + b_{L-1}).
\end{cases}
\end{equation}

Given the trained parameters and input, the output can be computed via a feed-forward process, which is used to evaluate whether regional frequency security is maintained or not.

\subsubsection{Formulating ICNN as Constraints}
To integrate the trained ICNN into the regional frequency-constrained model, the big $M$ method based on binary variable $\varsigma_{i}$ can be adopted to encode the ICNN in \eqref{eq:icnn structure} as the following constraints:
\vspace{-0.18em}
\begin{equation}\label{eq:icnn milp}
\begin{cases}
z_{i+1} \geq z_{i}^{T}W_{i} + s^{T}F_{i} + b_{i}, \\
z_{i+1} \leq z_{i}^{T}W_{i} + s^{T}F_{i} + b_{i} + M(1-\varsigma_{i}), \\
z_{i+1} \geq 0, 
z_{i+1} \leq M \varsigma_{i},
z_{L} \geq RFI.
\end{cases}
\end{equation}
where $RFI$ represents the indicator for regional frequency security. Given certain input $s$, the output of the final layer (i.e., $z_{L}$ in \eqref{eq:icnn milp}) should be larger or equal to $RFI$ (e.g., 0.5) for regional frequency security. In this case, frequency security can be ensured by including \eqref{eq:icnn milp} into the original optimisation model instead of the differential equations in \eqref{eq:coupled swing}-\eqref{eq:stiffness}.

\subsubsection{Limitations of Conventional ICNNs}
To effectively apply ICNNs to the regional frequency-constrained optimisation problem, two limitations should be appropriately addressed: 1) due to the nature of non-negative weight matrices and vanishing gradients, the convergence of conventional ICNNs can be notably slow, eventually leading to poor training performance. In this case, the extracted constraints can also be considerably conservative, which may lead to much higher operation costs and investment costs. 2) non-negative weights cause a reduced fitting ability; thus, the information from the previous layer cannot be efficiently passed to the next layer. To make up for this and enhance the representational power, skip-connections between the input layer and hidden layers have been added to conventional ICNNs. However, including these `passthrough' links further complicates the network architecture and training process as well as increases the model complexity.

\subsection{Enhanced ICNNs with Principled Weight Initialisation}
\begin{figure}[t!]
\centering
{\includegraphics[width=0.5\textwidth]{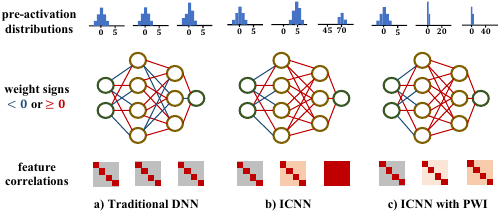}}
\vspace{-1.08em}
\caption{Illustration of the effects due to different signal propagation in hidden layers \cite{hoedt2024principled}. The top row shows histograms of pre-activations in each hidden layer. The bottom row displays the feature correlation matrices for these layers. Blue and red connections depict negative and positive weights, respectively.}
\vspace{-0.68em}
\label{fig:initialization}
\end{figure}

The reason for the poor convergence ability of traditional ICNNs is the bad initialisation of the non-negative weights. According to traditional signal propagation theory \cite{klambauer2017self}, it is normally assumed that the weights are sampled from a centred distribution, e.g., $\mathbb{E}[w_{ij}]\!=\!0$ and $Var[w_{ij}]\!=\!\sigma^2_{w}$, where $\mathbb{E}[w_{ij}]$ and $Var[w_{ij}]$ represent the mean and variance of the weight $w_{ij}$, respectively. However, this assumption is impractical for ICNNs, since weights are constrained to be non-negative \cite{hoedt2024principled}. As depicted in Fig. \ref{fig:initialization}, this poor initialisation can lead to distribution shifts that make training harder \cite{klambauer2017self}. To address this issue, a generalised signal propagation theory together with a novel principled weight initialisation (PWI) strategy is introduced in this section for non-negative weights of ICNNs, which can improve the training performance of ICNNs and eventually achieve similar training and test accuracy to traditional DNNs.

\subsubsection{Generalised Signal Propagation}
As depicted in Fig. \ref{fig:icnn}, ICNNs can be constructed by fully-connected networks, e.g., $f\!:\!\mathbb{R}^N \!\rightarrow\! \mathbb{R}^M$, with activation function $\phi: \mathbb{R} \rightarrow \mathbb{R}$ for element-wise vectors. The propagation of pre-activations throughout the neural network can be generalised as
\begin{equation}
s = W\phi(s^-) + b,
\end{equation}
where $s^{-}$ indicates pre-activations from the preceding layer.

Compared with traditional signal propagation theory, the weights should no longer be drawn from a zero-mean distribution, e.g., $\mathbb{E}[w_{ij}]\!=\! \mu_w \!\neq\! 0$, to accurately describe the effects caused by non-centred weights. Additionally, the effects of bias parameters are included for better control of signal propagation. Similar to weights, bias parameters are assumed to be $i.i.d.$ samples from the distribution with mean $\mathbb{E}[b_i] = \mu_b$ and variance $Var[b_i]=\sigma_b^2$ \cite{hoedt2024principled}. Thus, the first two moments of pre-activations can be defined as
\begin{equation}\label{first_moment}
\mathbb{E}[s_i] = N \mu_{w} \mathbb{E}[\phi(s^{-}_{1})] + \mu_b,
\end{equation}
\begin{equation}\label{second_moment}
\begin{split}
\mathbb{E}[s_{i}s_{j}] &= \delta_{ij}(N\sigma^{2}_{w}\mathbb{E}[\phi(s^{-}_{1})^2]+\sigma^{2}_{b})\\
&+\mu^{2}_{w}\sum_{k,k^{'}}Cov[\phi(s^{-}_{k}),\phi(s^{-}_{k^{'}})]+\mathbb{E}[s_{i}]\mathbb{E}[s_{j}],
\end{split}
\end{equation}
which depends on the variance propagation through the activation function $\phi(\cdot)$. Since mean and variance are independent of index $i$, it can be assumed that $\mathbb{E}[\phi(s^{-}_{k})^n]=\mathbb{E}[\phi(s^{-}_{1})^n],\forall k$. $\delta_{ij}$ is the Kronecker delta, while $Cov[\phi(s^{-}_{k}),\phi(s^{-}_{k^{'}})]$ refers to the off-diagonal elements of the covariance matrix, allowing to consider the covariance propagation through activations.

Due to $\mu_w \!\neq\! 0$, the pre-activations no longer have zero means by default. In addition, covariance plays a significant role in signal propagation. For example, the covariance propagation through the \texttt{ReLU} function \cite{daniely2016toward} can be specified as 
\begin{equation}\label{relu_activate}
\!\!\!\!\mathbb{E}[\texttt{ReLU}(s_1), \texttt{ReLU}(s_2)]\!=\!\frac{\sigma^2(\sqrt{1\!-\!\rho^2}\!+\!\rho \arccos(-\rho))}{2\pi},
\end{equation}
where $(s_1,s_2) \!\!\sim \!\!\mathcal{N}((0,0),\left(    \!\!
  \begin{array}{cc}   
  \small
    1 \!\!&\!\!\rho \\  
    \rho \!\!&\!\! 1 \\  
  \end{array} \!\!
\right) \sigma^2)$ and $\rho$ is the correlation. 

\subsubsection{Co-Variance Fixed Points}
To stabilise mean and covariance on top of the variance propagation and set up the fixed-point equation for the second moment in \eqref{second_moment}, we need to propagate through the \texttt{ReLU} non-linearity. However, the analysis of covariance propagation via activation functions (e.g., $\texttt{ReLU}$ in \eqref{relu_activate}) only holds if pre-activations have zero mean, which requires further modifications of the first moment in \eqref{first_moment}. Since we cannot enforce $\mu_w=0$, the bias parameters are modified to obtain centred pre-activations:
\vspace{-0.18em}
\begin{equation}
\mu_{b}=-N\mu_{w}\mathbb{E}[\phi(s^{-}_{1})],
\end{equation}
which allows referring to the second moment as (co-)variance.

Then, the propagation of the second moment in \eqref{second_moment} includes two parts. Given identically distributed pre-activations, first-part dynamics for off-diagonal entries can be defined as
\begin{equation}
\begin{split}
&Cov_{i\neq j}[s_{i},s_{j}]=\mu^{2}_{w}\sum_{k,k^{'}}Cov[\phi(s^{-}_{k}),\phi(s^{-}_{k^{'}})]\\
&=N\mu^{2}_{w}(Var[\phi(s^{-}_{1})]+(N-1)Cov[\phi(s^{-}_{1}),\phi(s^{-}_{2})]).
\end{split}
\end{equation}

Similarly, the second part models the on-diagonal entries \cite{hoedt2024principled}, i.e., the variance, which can be defined as
\begin{equation}
Var[s_{i}]=N\mu^{2}_{w}\mathbb{E}[\phi(s^{-}_{1})]+\sigma^2_{b}+Cov[s_1,s_2].
\end{equation}

In this context, the \texttt{ReLU} moments in \eqref{relu_activate} can be plugged into the above results \cite{hoedt2024principled} and the fixed-point equations in terms of correlation and variance can be obtained as follows:
\begin{equation}\label{variance}
\begin{cases}
\rho_* = \mu_w^2(\frac{N(\pi-N+(N-1)\sqrt{1 - (\rho^*)^2} + \rho^*\arccos(-\rho^*))}{2\pi}),\\
(\sigma^*)^2 = N\sigma^{2}_{w}\frac{1}{2}(\sigma^*)^2+\sigma^2_{b}+\rho^*(\sigma^*)^2,
\end{cases}
\end{equation}
where $\rho^*=\frac{1}{(\sigma^*)^2}Cov[s^{-}_{1},s^{-}_{2}]=\frac{1}{(\sigma^*)^2}Cov_{i \neq j}[s_{i},s_{j}]$ and $(\sigma^*)^2=Var[s^{-}_{1}]=Var[s_i]$.

\subsubsection{Weight Distribution for ICNNs}
The fixed point equations in \eqref{variance} can be directly solved to obtain distribution parameters for initial weights of ICNNs, e.g., $\sigma^2_{b}$, $\sigma^2_{w}$, and $\mu^2_{w}$. By setting $\sigma^2_{b}=0$, the initialisation parameters for ICNNs can be derived as follows:
\begin{equation}
\!\!
\begin{cases}
\mu_w^2 \!=\! \frac{2\pi\rho_*\left(\pi - N + (N-1)\left(\sqrt{1-\rho_*^2} + \rho_*\arccos(-\rho_*)\right)\right)^{-1}}{N},\\
\sigma^2_{w}=\frac{2}{N}(1-\rho_*),
\end{cases}
\end{equation}
which only depend on correlation $\rho_{*}$ rather than variance $\sigma^2_{*}$. Following the same practice in \cite{hoedt2024principled}, $\rho_{*}\!=\!1/2$ is selected as a compromise to obtain the following initialisation parameters:
\begin{equation}
\begin{cases}
    \mu_{w} = \sqrt{\frac{6\pi}{N(6(\pi - 1) + (N - 1)(3\sqrt{3} + 2\pi - 6))}},~~
\sigma_{w}^{2} = \frac{1}{N},\\
\mu_{b} = \sqrt{\frac{3N}{6(\pi - 1) + (N - 1)(3\sqrt{3} + 2\pi - 6)}},~~
\sigma_{b}^{2} = 0.
\end{cases}
\end{equation}

For non-negative weights in ICNNs, log-normal distribution can be used for sampling, which is expressed as
\begin{equation} \label{eq:sample}
\begin{cases}
\mu_{w}' = \ln(\mu_{w}^{2}) - \frac{1}{2} \ln(\sigma_{w}^{2} + \mu_{w}^{2}),\\
\sigma_{w}'^{2} = \ln(\sigma_{w}^{2} + \mu_{w}^{2}) - \ln(\mu_{w}^{2}),
\end{cases}
\end{equation}
where \eqref{eq:sample} ensures that sampled weights are non-negative and have the desired mean and variance. To this end, due to the introduction of PWI strategy, the skip-connections $F$ between the input layer and hidden layers can be omitted, leading to a much more concise constraint setup of ICNNs as follows:
\begin{equation}\label{eq:icnn no}
\begin{cases}
z_{i+1} \geq z_{i}^{T}W_{i} + b_{i}, \\
z_{i+1} \leq z_{i}^{T}W_{i} + b_{i} + M(1-\varsigma_{i}), \\
z_{i+1} \geq 0, 
z_{i+1} \leq M \varsigma_{i},
z_{L} \geq RFI.
\end{cases}
\end{equation}
\vspace{-0.25em}

\section{Reformulation of the Regional Frequency-Constrained Planning Model}
\label{sec:IV}
\subsection{Final Mathematical Formulation}
Using the above extracted regional frequency security constraint \eqref{eq:icnn no}, the final mathematical formulation of the proposed planning model can be expressed as follows:
\begin{equation}\label{eq:model}
\begin{split}
obj.~
\min_{G^{cap},RES^{cap},ES^{cap}}~\eqref{eq:obj}~~~~~~~~~~~~~~~~\\
    s.t.~\begin{cases}\textit{generator constraints:}~\eqref{eq:gen energy range}-\eqref{eq:gen mut},\\
    \textit{ES and RES constraints}~\eqref{eq:ev model charge}-\eqref{acpowerres},\\
    \textit{power flow constraints:}~\eqref{eq:acopf_bal1}-\eqref{eq:acopf_thermal},\\
    \textit{carbon constraint:}~\eqref{eq:carbon},\\
    \textit{frequency constraints:}~\eqref{eq:tso qss}\!-\!\eqref{eq:ev efr_limit_2},\eqref{eq:icnn no}.
    \end{cases}
\end{split}
\end{equation}

\subsubsection{Reformulation of Conditional Constraints}
\label{sec:IV.A}
Equations \eqref{eq:gen ramp up range} and \eqref{eq:gen ramp down range} belong to conditional constraints, which can be embedded to the optimisation by introducing auxiliary binary variables \cite{sioshansi2017optimization}. Take \eqref{eq:gen ramp up range} as an example, if the generator $g$ has three ramp-up rates ($R_1$, $R_2$, and $R_3$) and two elbows ($E_1$ and $E_2$), conditional constraint \eqref{eq:gen ramp up range} can be rewritten as
\begin{equation}\label{eq:condition}
\begin{split}
\begin{cases}
     P_{g,t}^{w} \!- P_{g,t-1}^{w} \leq R_1 \cdot y^{w}_{g,t}, ~if~P_{g,t-1}^{w} \in [0,E_1),\\
     P_{g,t}^{w} \!- P_{g,t-1}^{w} \leq R_2 \cdot y^{w}_{g,t}, ~if~P_{g,t-1}^{w} \in [E_1,E_2),\\
     P_{g,t}^{w} \!- P_{g,t-1}^{w} \leq R_3 \cdot y^{w}_{g,t}, ~if~P_{g,t-1}^{w} \in [E_2,\overline{P}_{g,t}^{w}],
\end{cases}
\end{split}
\end{equation}
where $E_1 < E_2 < \overline{P}_{g,t}^{w}$. To embed the conditional constraint \eqref{eq:condition} into the optimisation, binary variables $z^{1,w}_{g,t-1}$, $z^{2,w}_{g,t-1}$, $z^{3,w}_{g,t-1}$, and $z^{4,w}_{g,t-1}$ are introduced to reformulate \eqref{eq:condition} as
\begin{equation}\label{eq:condition solve}
\begin{split}
\!\!
\begin{cases}
     P_{g,t-1}^{w} \!+M z^{1,w}_{g,t-1} \geq E_1,\\
     P_{g,t}^{w} \!- P_{g,t-1}^{w}-M (1-z^{1,w}_{g,t-1}) \leq  y^{w}_{g,t}R_1,\\
     P_{g,t-1}^{w} \!-M z^{2,w}_{g,t-1} \leq E_1-m,\\ 
     P_{g,t}^{w} \!-\! P_{g,t-1}^{w}\!-\!M (1\!-\!z^{2,w}_{g,t-1})
     \!-\!M z^{3,w}_{g,t-1} \!\leq\! y^{w}_{g,t}R_2, \\
      P_{g,t-1}^{w} \!-M z^{3,w}_{g,t-1} \leq E_2-m,\\ 
     P_{g,t}^{w} \!- P_{g,t-1}^{w}-M (1-z^{3,w}_{g,t-1}) \leq y^{w}_{g,t}R_3, \\
     P_{g,t-1}^{w} \!+M z^{4,w}_{g,t-1} \geq E_2,\\ 
     z^{2,w}_{2,g,t-1}+z^{3,w}_{g,t-1} - M (1-z^{4,w}_{g,t-1}) \leq 1,
\end{cases}      
\end{split}
\end{equation}
where $M$ refers to a very big number (e.g., $10^{5}$), while $m$ is a very small number (e.g., $10^{-5}$) to avoid numerical issues. Note that the employed big-M method in constraint \eqref{eq:condition solve} has been well studied in existing literature \cite{badesa2020pricing}, which can ensure the uniqueness of the generator's ramp-up and ramp-down rates at a certain time step. Specifically, when the $M$ is sufficiently big (e.g., $10^{5}$) and $m$ is small enough (e.g., $10^{-5}$) with respect to decision variables in the model, the solution quality can be guaranteed. The ramping rate used at time $t$ only depends on the generation output level at time $t-1$. For example, when the generation output at time $t-1$ is within the range of $[0,E_1)$, only the ramping rate $R_1$ should be used in the ramping process from time $t-1$ to time $t$.
\vspace{-0.48em}

\subsection{Solving Methodology}
\label{sec:IV.C}
As mentioned in the second challenge of Section \ref{sec:II.C}, the final mathematical model \eqref{eq:model} is highly complicated, due to the incorporation of numerous integer variables, especially for investment decisions for optimal sizing, UC setup of generators based on conditional constraints, and the integration of an ICNN. In particular, when large-scale power systems are considered, the obtained mathematical model will easily become computationally intractable. 

In this section, an adaptive genetic algorithm (AGA) with sparsity calculation and local search is developed to further divide the final mathematical model \eqref{eq:model} into two levels by separately handling integer variables relating to invested capacities and locations (upper-level) and frequency-constrained operation decision-making based on the enhanced ICNN (low-level). Furthermore, an additional module of dynamic simulation is integrated into the operation level for the verification of regional frequency security, ensuring the safety of ICNN-based frequency constraints. In detail, the proposed AGA is first used to generate a number of initial populations including different optimal sizing decisions and simulate the planning level in \eqref{eq:obj}. Afterwards, the generated scenarios will be sent to the operation level with the objective function in \eqref{eq:tso obj} to minimise operation costs under different sizing decisions. Note that the sum of planning costs and operation costs will be used as the fitness function to evaluate the superiority of these populations, followed by two genetic operators (crossover and mutation) that alter the composition of the children. Then, sparsity calculation and local search operators are introduced to further explore the possibility of better solutions, where the obtained new children will also be evaluated by the fitness function based on the regional frequency-constrained planning model \eqref{eq:model}. In this manner, the proposed planning model can be effectively separated into planning and operation levels, and then iteratively solved by the combination of the proposed AGA and commercial software such as Gurobi. The process of the enhanced AGA is depicted in Fig. \ref{fig:aga}, while algorithm details are given below:

\subsubsection{Adaptive Genetic Algorithm}
GA is categorised as a global search meta-heuristic that does not suffer from parameter numbers and is efficient in terms of computational time and programming simplicity \cite{wang2021three}. However, traditional GA, which normally adopts a small mutation rate (e.g., 0.01-0.1), may easily fall within the local minimum, while a large mutation rate (e.g., over 0.3) can make the problem hard to converge. In this context, this paper introduces AGA with adaptive crossover and mutation probabilities to avoid local optimum. The crossover and mutation probabilities ($p^c$ and $p^m$) of the suggested AGA can be calculated as
\begin{equation} \label{adaptive}
\begin{cases}
p^c = k_1 \frac{F^{max} - F^{ave}}{F^{max}-F'},~when~F' \leq F^{ave},\\
p^c = k_3,~when~F' > F^{ave},\\
p^m = k_2 \frac{F^{max} - F^{ave}}{F^{max}-F'},~when~F' \leq F^{ave},\\
p^m = k_4,~when~F' > F^{ave},
\end{cases}
\end{equation}
where $F^{max}$, $F^{ave}$, and $F'$ refer to the largest fitness value in the operation of crossover, the average fitness value, and the smaller fitness value within the population, respectively. $k_1$, $k_2$, $k_3$, $k_4$ are constants, where the choice of their values is empirical and can be changed in different problems \cite{wang2021three}.

\begin{figure}
\centering
\includegraphics[width=0.46\textwidth]{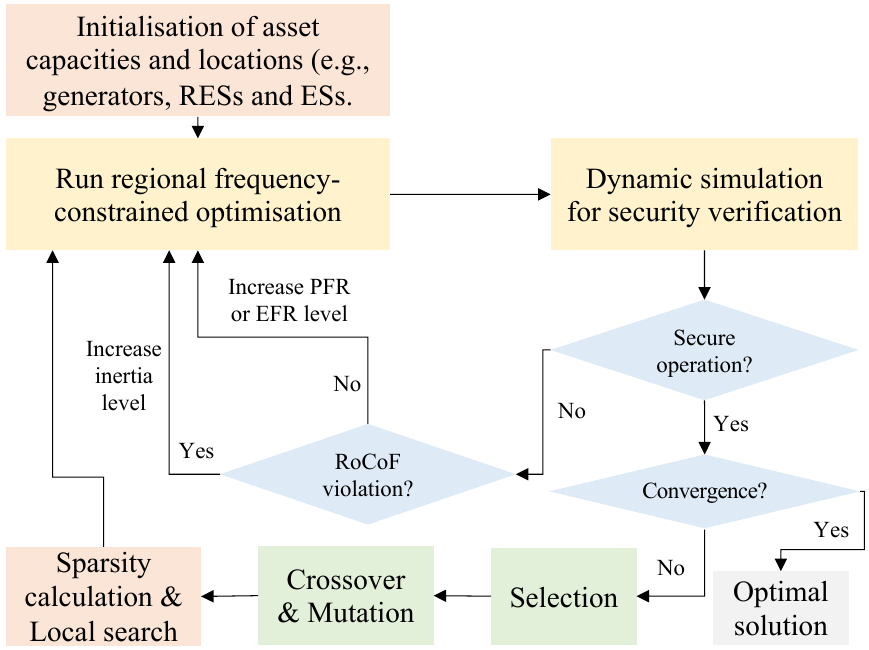}
\caption{Illustration of the proposed enhanced AGA with adaptive crossover and mutation probabilities as well as local search based on sparsity calculation.}\label{fig:aga}
\vspace{-0.88em}
\end{figure}

\subsubsection{Sparsity Calculation and Local Search}
To enhance the diversity of the GA population, sparseness theory and local search \cite{luan2023enhanced} are incorporated into the AGA, which allows the algorithm to obtain new children populations by executing local search operations on the sparse solution. Thus, both population diversity and algorithm generality can be improved.

In detail, we first compute the Euclidean distance between one individual and other individuals, which is written as
\begin{equation}
    \rho = \sqrt{\sum_{i=1}^{J}(x_{i}-y_{i})^2},
\end{equation}
where $J$ is the population size. Then, the sparsity of the individual $x_{i}$ can be expressed as
\begin{equation}
    SP(x_i)=U_i/J,
\end{equation}
where $U_i$ represents the number of individuals whose Euclidean distance from other individuals is less than the judgment threshold \cite{luan2023enhanced}. As a result, the individual with the smallest sparsity $SP(x_i)$ can be defined as the sparse solution.

After the sparse solution is determined, the children population of local search can be obtained by performing the following two local search operators $O_1$ and $O_2$:
\begin{itemize}
\item[$O_1$:] Randomly select two elements from the sparse solution and conduct position exchange on them.
\item[$O_2$:] Randomly select one element from the sparse solution and reduce its value by $c$ (e.g., 2).
\end{itemize}

In this context, sparsity calculation plays a crucial role in guiding the AGA by identifying areas within the solution space that exhibit sparse characteristics. Once the sparsity pattern is identified, the AGA performs a more targeted local search around the sparse solution. This focused approach enables the algorithm to delve deeper into areas where potential improvements are most likely to be found, thereby enhancing the efficiency of the local search process and eventually leading to faster convergence and more accurate optimisation results.

\subsubsection{AGA Population Design}
As shown in Fig. \ref{fig:aga}, capacities and locations of generators, RESs, and ESs for the upper-level problem can be initialised as iterated generations (e.g. pairs of generator ratings [200 MW, 300 MW] within the range of [0 MW, 350 MW]) in the AGA. The fitness function is the whole system cost \eqref{eq:obj} provided iteratively for each chromosome through regional frequency-constrained optimisation. To ensure reliable planning solutions, AGA is considered to have converged, only if the obtained solution remains unchanged for a specified number of iterations (e.g., 20 or 30) or the maximum allowable number of iterations is reached (e.g., 100). Furthermore, the integrated dynamic simulation module is primarily used for the security verification of frequency ancillary service schedules. If regional frequency can be secured, the AGA iterative process continues; otherwise, a lower bound will be generated by slightly increasing the current service level and then added to the regional frequency model for iteration until secure operation is achieved. It is worth noting that this security verification process is only used to ensure the obtained solutions from the proposed regional frequency model meet the requirements of frequency security, following a similar practice in \cite{nakiganda2022stochastic}. However, the primary focus of the proposed AGA algorithm is still to solve the regional frequency-constrained planning model via different genetic operators and obtain cost-effective investment decisions considering frequency security.
\vspace{-0.08em}

\section{Case Studies}
\vspace{-0.08em}
\label{sec:V}
\subsection{Experimental Setup}
\vspace{-0.00em}
\label{sec:V.A}
\subsubsection{Data Descriptions}
\label{sec:V.A.1}
\begin{figure}[h!]
\vspace{-0.08em}
\centering
{\includegraphics[width=0.48\textwidth]{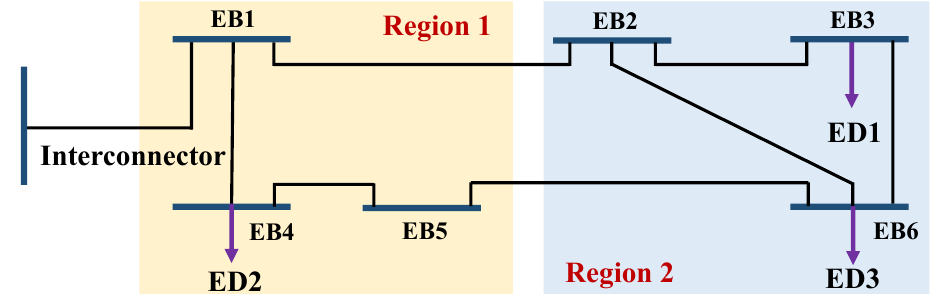}}
\vspace{-0.08em}
\caption{The modified IEEE 6-bus power network with one interconnector.}
\label{fig:network}
\vspace{-0.28em}
\end{figure}

Case studies are conducted on a modified IEEE 6-bus power system, a 14-bus GB power system, and a modified IEEE 118-bus power system, where the network structure of the 6-bus system is illustrated in Fig. \ref{fig:network}. The yearly dataset used to generate representative days is collected from \cite{nationaldata}. Cost and operation data for different types of generation technologies including CCGT, OCGT, NG CCS, WTs, and PVs are collected from \cite{zhang2018whole,badesa2019simultaneous}, while the cost and operation data of ESs can be found in \cite{wang2021three}. The carbon target is set as 100 g/kWh \cite{zhang2018whole}.

\subsubsection{Frequency Security Setup}
\label{sec:V.A.2}
Regarding post-fault frequency limits, the frequency nadir should never be lower than 49.2 Hz to avoid activating under-frequency load shedding, i.e., $\Delta \overline{f}=0.8~Hz$ \cite{badesa2019simultaneous}. Since the RoCoF limit is being relaxed to 1 Hz/s in the GB power system \cite{sqss}, the RoCoF should be below 1 Hz/s to prevent the tripping of RoCoF-sensitive protection relays, i.e., $\overline{RoCoF}=1~Hz/s$. The delivery speed of frequency response from ESs and generators is set as 1 s and 10 s, respectively \cite{badesa2019simultaneous}. 

\subsubsection{ICNN Network Setup}
\label{sec:V.A.3}
To learn regional frequency security, an ICNN with two hidden layers is built. The number of neurons is 64 and 32, respectively. The input of the ICNN consists of inertia, PFR, EFR, demand, fault information (i.e., largest loss and fault location), etc. The output is the indicator for regional frequency security. The learning rate of the Adam optimiser is 0.001. The batch size is set as 64, while the convergence threshold is set as 0.1\%. The maximum iteration and update interval are set as 1000 and 10, respectively. Furthermore, 15,000 samples are randomly generated via dynamic simulations in MATLAB/Simulink, which are then divided into training sets, validation sets, and test sets with a ratio of 7:1.5:1.5, following a similar practice in \cite{wu2023transient}.
\vspace{-0.68em}

\subsection{Performance of the Proposed  Enhanced ICNN}
\begin{figure}[t!]
\vspace{-0.08em}
\centering
{\includegraphics[width=0.493\textwidth]{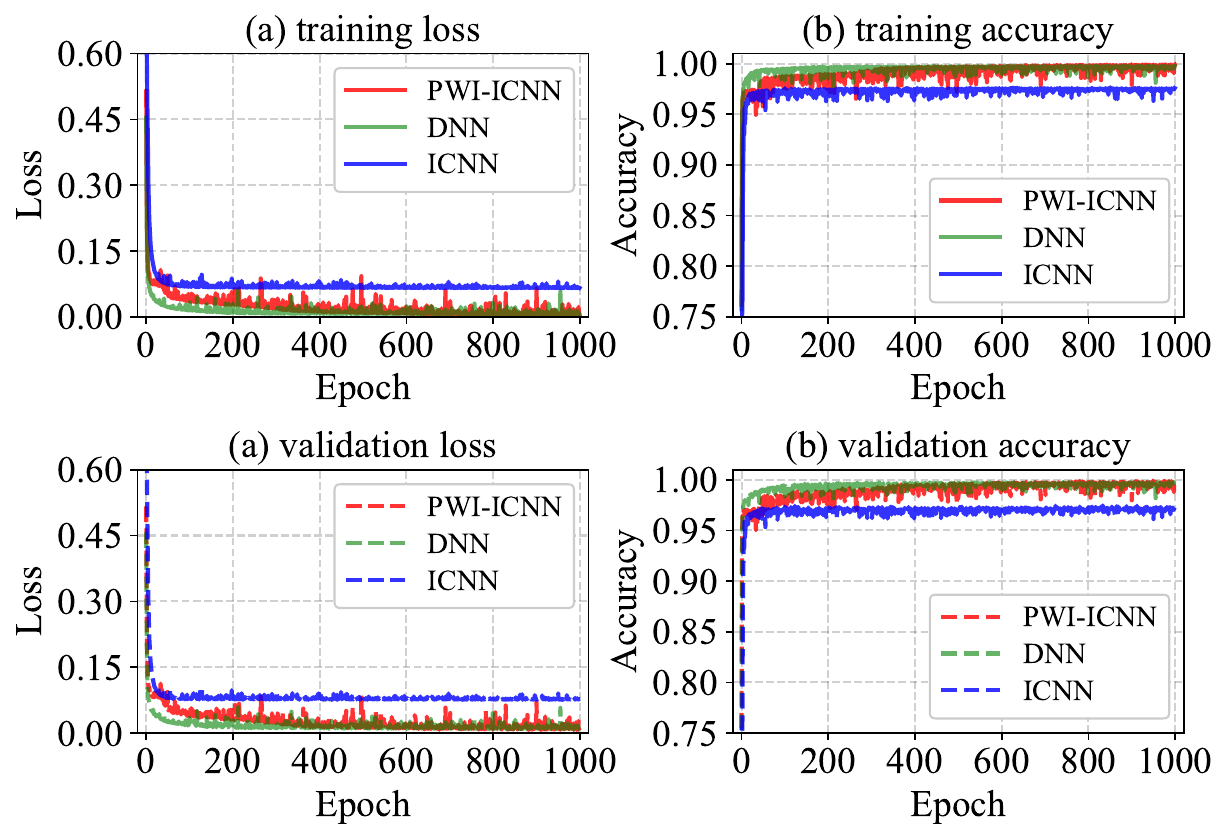}}
\vspace{-1.88em}
\caption{Losses and accuracy of training and validation for the proposed ICNN-based model with PWI strategy, the DNN-based model, and traditional ICNN-based model.}
\label{fig:train_test}
\vspace{-1.40em}
\end{figure}

\vspace{-0.08em}
\label{sec:V.B}
To verify the performance of the proposed enhanced ICNN model on generating effective regional frequency constraints, a comparison has been conducted, including three different models, i.e., the DNN-based model, the traditional ICNN-based model, and the proposed ICNN model based on the PWI strategy. The training and validation performance including the evolution of the loss function and the accuracy is demonstrated in Fig. \ref{fig:train_test}. In addition, the performance comparison regarding the test dataset
is summarised in Table \ref{table:f1}, where metrics include `precision', `recall', and `F1-score'.

It can be observed in Fig. \ref{fig:train_test} (a)-(b) that the proposed PWI-ICNN model obtains lower training loss and higher training accuracy (around 99.0\%), compared to the traditional ICNN model with a training accuracy of around 97.5\%. To prevent overfitting, the validation loss and accuracy are also recorded in Fig. \ref{fig:train_test} (c)-(d), which demonstrates a similar conclusion, i.e., the proposed PWI-ICNN model receives much lower validation loss and higher validation accuracy. The reason is that the proposed PWI-ICNN adopts the novel principled weight initialisation strategy, which can effectively deal with non-decreasing convex activation functions and non-negative weights of traditional ICNNs, effectively accelerating DNN learning and eventually leading to better generalisation performance. In addition, it can be observed from Fig. \ref{fig:train_test} that the proposed PWI-ICNN model eventually exhibits similar learning behaviour as the non-convex DNN baseline, due to the integration of PWI strategy. As a result, the PWI-ICNN model can match the performance of non-convex DNNs in terms of accuracy and learning dynamics, while still preserving the key advantages of convexity, such as improved tractability and reliability in optimisation.

Regarding the test performance, it can be observed from Table \ref{table:f1} that the proposed PWI-ICNN model has an F1-score of 1.00 for unsafe operation points and 0.99 for safe operation points, while the traditional ICNN model has F1-scores of 0.98 and 0.95 for unsafe operation points and safe operation points, respectively. Therefore, it can be concluded that the proposed PWI-ICNN model exhibits superior performance across all metrics compared to the traditional ICNN model, particularly excelling in the `precision' and `recall' of both classes (unsafe and safe). The convexity and performance of the final optimisation model can both be ensured.
\vspace{-0.28em}

\begin{table}[h!]
\vspace{-0.68em}
\footnotesize
\centering
\renewcommand\arraystretch{1.00}
\setlength{\abovecaptionskip}{12pt}
\caption{Comparison of test performance among PWI-ICNN, ICNN and DNN.}
\setlength{\tabcolsep}{2.68mm}{
\begin{tabular}{|c|c|c|c|c|}
\toprule
& Class & Precision & Recall & F1-Score  \\ \midrule
\multirow{2}{*}{PWI-ICNN}& unsafe   & 1.00      & 1.00   & 1.00         \\
& safe  & 1.00      & 0.99   & 0.99         \\ \midrule
\multirow{2}{*}{ICNN}& unsafe   & 0.98      & 0.98   & 0.98         \\
& safe  & 0.94      & 0.95   & 0.95   \\ \midrule
\multirow{2}{*}{DNN}& unsafe   & 1      & 1.00   & 1.00         \\
& safe  & 0.99      & 1   & 0.99 \\ \bottomrule
\end{tabular}}
\label{table:f1}
\vspace{-1.08em}
\end{table}

\subsection{Analysis of the Planning Results}
\begin{figure}[h!]
\vspace{-0.38em}
\centering
{\includegraphics[width=0.49\textwidth]{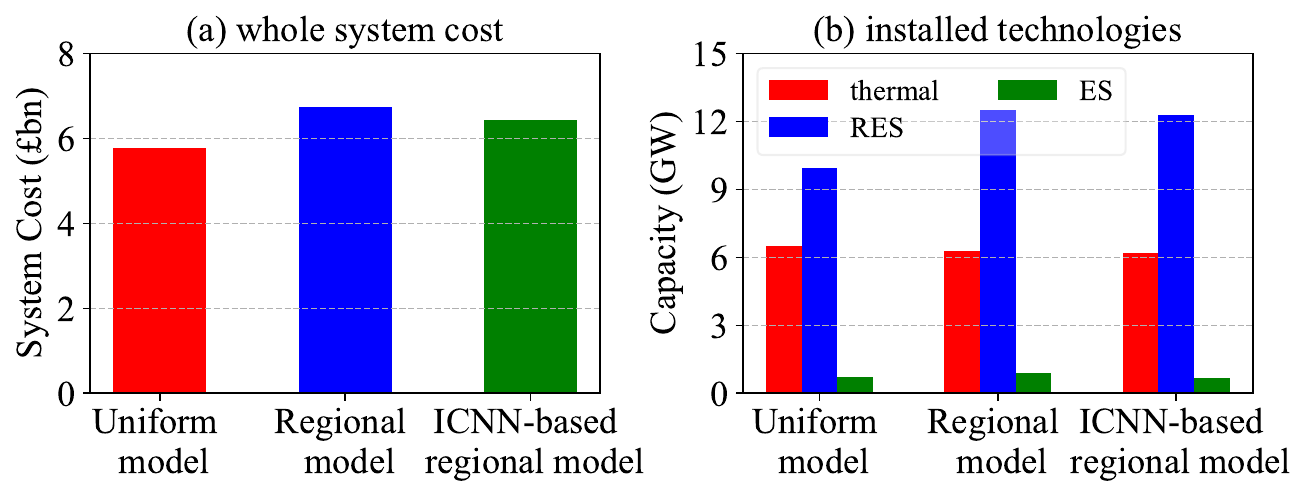}}
\vspace{-1.68em}
\caption{Comparisons among 3 different cases concerning the whole system cost as well as installed capacities of conventional generation, RES, and ESs.}
\label{fig:cost}
\vspace{-0.28em}
\end{figure}

\begin{figure}[h!]
\vspace{-0.38em}
\centering
{\includegraphics[width=0.49\textwidth]{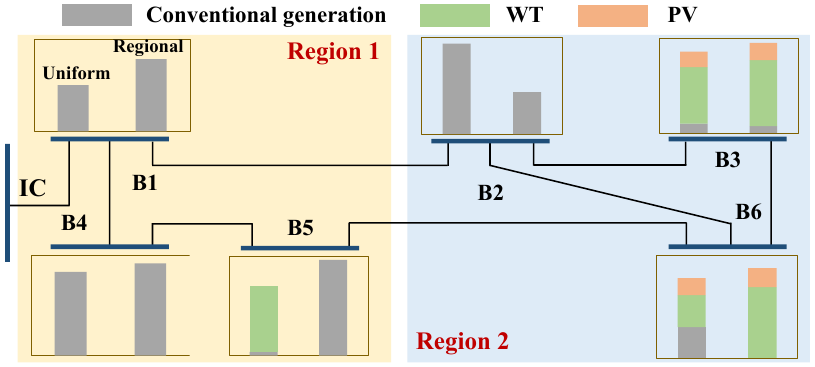}}
\vspace{-1.48em}
\caption{Comparison of resource capacity and allocation between the uniform frequency model and the proposed ICNN-based regional frequency model.}
\label{fig:network_com}
\vspace{-0.18em}
\end{figure}

\begin{figure}[h!]
\vspace{-0.38em}
\centering
{\includegraphics[width=0.49\textwidth]{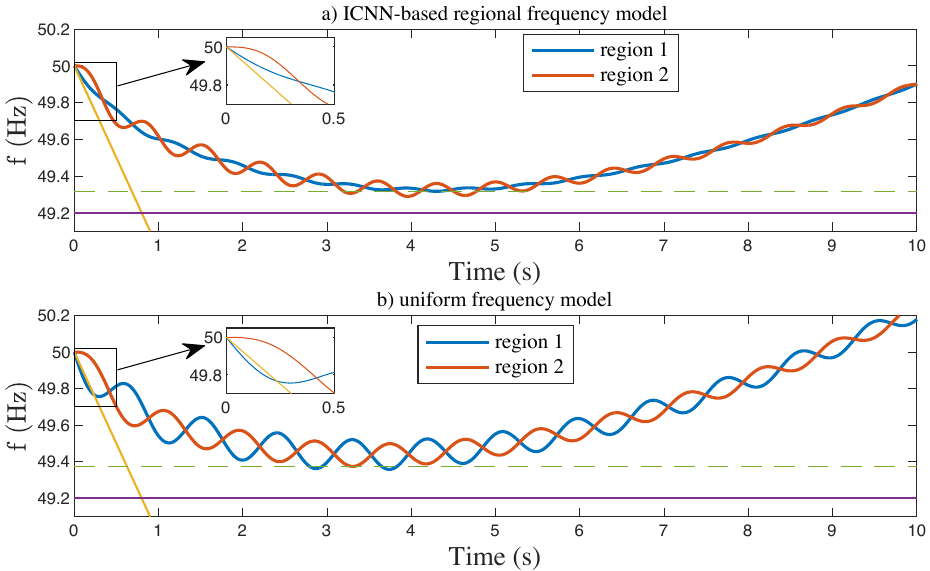}}
\vspace{-1.58em}
\caption{Results of frequency deviations under uniform frequency model and ICNN-based regional frequency model concerning the same power infeed loss.}
\label{fig:matlab}
\vspace{-0.38em}
\end{figure}

\begin{figure}[t!]
\vspace{-0.08em}
\centering
{\includegraphics[width=0.495\textwidth]{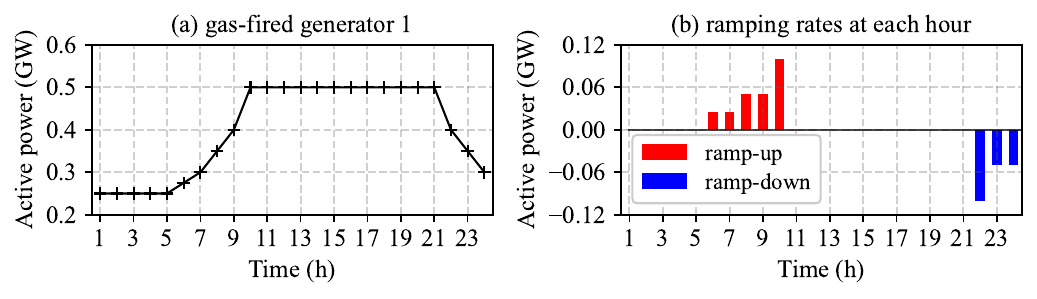}}
\vspace{-2.18em}
\caption{Illustration of the scheduling behaviours of one conventional gas-fired generator with three different ramping rates and two elbows.}
\label{fig:uc}
\vspace{-1.28em}
\end{figure}

\vspace{-0.08em}
\label{sec:V.C}
To demonstrate the effectiveness and benefits of the proposed regional frequency-constrained planning model in accurate and reliable decision-making, three different cases are prepared for a detailed comparison of investment results: 1) uniform frequency model developed in \cite{badesa2019simultaneous} that ignores inter-area frequency oscillations, 2) analytical-numerical regional frequency model developed in \cite{badesa2021conditions} where the regional frequency constraints are deduced in an analytical manner, 3) the proposed enhanced ICNN-based regional frequency model where the regional frequency constraints are formulated via the well-trained enhanced ICNN. The comparisons among the above three cases concerning the whole system costs and installed capacities of different technologies are illustrated in Fig. \ref{fig:cost}. The comparison between the proposed ICNN-based regional frequency model and uniform frequency model is conducted, where the difference in investment decisions is illustrated in Fig. \ref{fig:network_com} and the advantages/benefits of the proposed ICNN-based model in ensuring regional frequency security are further demonstrated in Fig. \ref{fig:matlab}. In addition, the detailed scheduling behaviours of one conventional gas-fired generator with three different ramping rates and two elbows are presented in Fig. \ref{fig:uc}.

\subsubsection{Investment results}
It can be observed from Fig. \ref{fig:cost}(a) that considering regional frequency security requirements (analytic regional model and ICNN-based regional model) can increase the whole system cost, compared with the uniform model which does not consider regional frequency security. The reason is twofold. On one hand, more frequency ancillary services are required under regional frequency models to provide sufficient security support and additionally deal with inter-area frequency oscillations, leading to a larger number of part-loaded generators and increasing operation costs.

On the other hand, it can be observed from Fig. \ref{fig:cost}(b) that RES penetration is increased under two `regional frequency' cases, compared with the uniform frequency model. This is because, when regional frequency security is considered, inertia and thermal generation have to be carefully allocated depending on specific regions rather than mainly considering the availability of natural resources, e.g., wind speed and solar irradiation. As demonstrated in Fig. \ref{fig:network_com}, thermal generation under the ICNN-based regional frequency model tends to be allocated at region 1, since this region is connected with the IC, easily leading to the potential largest power infeed/outfeed loss. In detail, much more thermal generation is allocated in buses 1, 4, and 5, while bus 5 is rich in renewable natural resources. To ensure demand-supply balance and meet the carbon target, more RESs have to be installed in region 2, e.g., bus 6 with relatively insufficient natural resources, eventually leading to larger RES capacity investment under regional frequency cases, compared with the uniform frequency model. 

Furthermore, it can be observed from Fig. \ref{fig:cost}(a) that the proposed ICNN-based regional frequency model obtains lower whole system costs compared with the analytical-numerical regional frequency model. This cost reduction is due to the ICNN-based model’s use of a broader dataset across diverse operating conditions, allowing it to provide a more general and less conservative solution. In contrast, the analytical-numerical model relies on frequency constraints derived from specific operating conditions, which cover a narrower range and result in over investment.

\begin{figure}[h!]
\vspace{-0.78em}
\centering
{\includegraphics[width=0.49\textwidth]{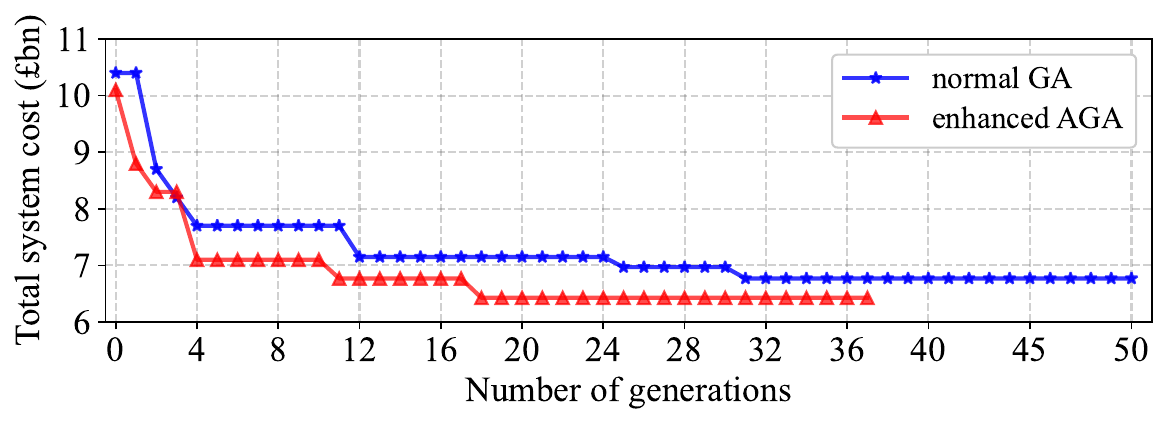}}
\vspace{-1.68em}
\caption{Total system cost evolution curves of the normal GA and the proposed AGA with sparsity calculation and local search.}
\label{fig:aga_result}
\vspace{-0.38em}
\end{figure}

Finally, to demonstrate the performance of the proposed enhanced AGA in avoiding local minimum and finding cost-effective solutions, a comparison has been conducted, including two cases: a) normal GA without adaptive probabilities and local search, and b) the proposed AGA with sparsity calculation and local search. To ensure solution quality, a strict convergence criterion is used to determine if the AGA has converged or not, i.e., the selected best solution does not change for over 20 iterations. It can be observed from Fig. \ref{fig:aga_result} that the proposed AGA reaches convergence after 38 iterations, which is faster than the normal GA with 51 iterations. Furthermore, the proposed AGA obtains a final solution with a lower system cost than the normal GA, exhibiting its effectiveness in avoiding local minimum.
\vspace{-0.08em}

\subsubsection{Operation results}
To analyse the performance of the proposed ICNN-based frequency model for post-fault frequency security, a comparison case based on dynamic simulations in MATLAB/Simulink is conducted including both ICNN-based regional frequency model and uniform frequency model. The operating points such as inertia, PFR and EFR obtained from these two models are fed into the dynamic simulation under the same power infeed loss 0.75 GW to simulate the post-fault frequency evolution curves. The accuracy of the proposed regional frequency-constrained model in ensuring regional frequency security is verified using two main criteria: 1) the RoCoF in each region must remain below 1 Hz/s, and 2) the frequency nadir in each region must exceed 49.2 Hz \cite{badesa2021conditions}. For regional frequency security, it is essential that both criteria are satisfied simultaneously across all regions.

As shown in Fig. \ref{fig:matlab}, under the same power infeed loss, the ICNN-based frequency model can successfully ensure regional frequency security including both frequency nadir and RoCoF, and lead to much less conservative solutions, compared with the uniform frequency model. In detail, there is a certain level of RoCoF violations in region 1 under the uniform frequency model, while the frequency nadir is at around 49.4 Hz, slightly higher than the frequency nadir (around 49.3 Hz) under the ICNN-based regional frequency model. The ICNN-based frequency model also receives much smaller inter-area frequency oscillations than the uniform frequency model. These results demonstrate the superior performance of the proposed regional frequency-constrained model in achieving solutions with security insurance and high quality. In addition, the detailed power schedules of one gas-fired generator are illustrated in Fig. \ref{fig:uc}. It can be found that the generator experiences three different ramp-up rates and two elbows during the selected representative day. The advantages of the proposed planning model in capturing detailed scheduling behaviours of each generator have been verified.

\begin{figure}[h!]
\vspace{-0.08em}
\centering
{\includegraphics[width=0.495\textwidth]{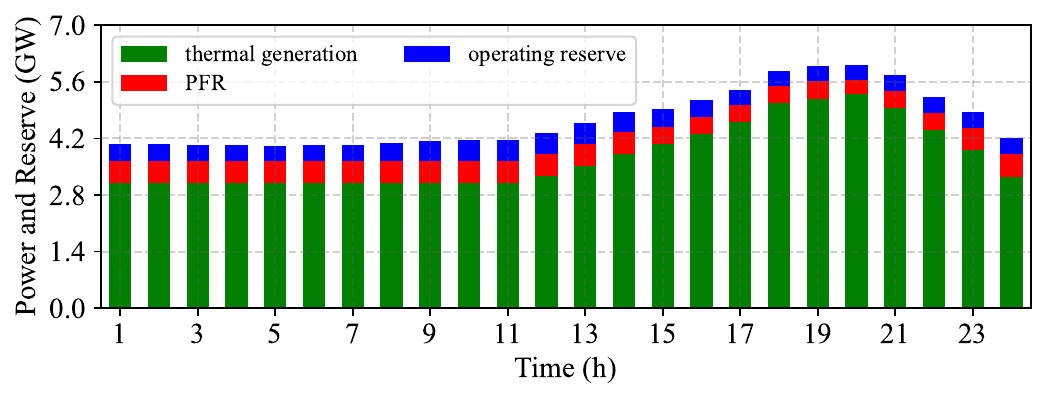}}
\vspace{-2.18em}
\caption{Illustration of the results of thermal generation, PFR, and operating reserve under one selected day.}
\label{fig:reserve}
\vspace{-0.88em}
\end{figure}

As indicated in reserve constraint \eqref{eq:reserve_new}, the unpredictability of substantial RES integration should be considered in the proposed optimal sizing model. To analyse its impact on system costs, a sensitivity analysis with varying RES forecasting errors has been carried out, with results summarised in Table \ref{table:forecast}. The analysis shows that greater RES unpredictability can increase total system costs, since more reserve needs to be procured. Additionally, the operational results for thermal generation, PFR, and operating reserve under a 10\% forecasting error are illustrated in Fig. \ref{fig:reserve}, where the operating reserve is scheduled to accommodate potential RES forecasting errors.

\begin{table}[h!]
\vspace{-0.28em}
\footnotesize
\centering
\renewcommand\arraystretch{1.00}
\setlength{\abovecaptionskip}{12pt}
\caption{Sensitivity analysis on the influence of forecasting errors concerning the whole system costs.}
\setlength{\tabcolsep}{2.28mm}{
\begin{tabular}{|c|c|c|}
\toprule
No. & Average Forecasting Error       & System Cost ($\pounds bn$)  \\ \midrule
1 & 5\%    & 6.40             \\ \midrule
2 & 10\%   & 6.43               \\ \midrule
3 & 15\%   & 6.46               \\ \midrule
4 & 20\%   & 6.50               \\ \midrule
5 & 25\%  & 6.52          \\ \bottomrule
\end{tabular}}
\label{table:forecast}
\end{table}

\vspace{-0.9em}

\subsection{Scalability Analysis in Larger Power Systems}
\begin{figure}[h!]
\vspace{-0.48em}
\centering
{\includegraphics[width=0.49\textwidth]{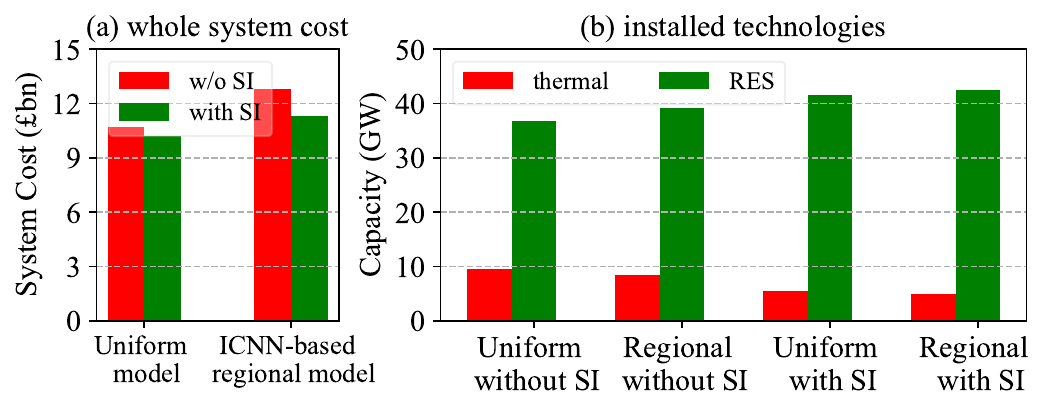}}
\vspace{-1.38em}
\caption{Comparisons between uniform frequency-constrained planning and regional frequency-constrained planning concerning the whole system cost as well as the installed thermal and RES capacities in the 14-bus power system.}
\label{fig:big_cost}
\vspace{-0.48em}
\end{figure}

\begin{figure}[h!]
\vspace{-0.48em}
\centering
{\includegraphics[width=0.49\textwidth]{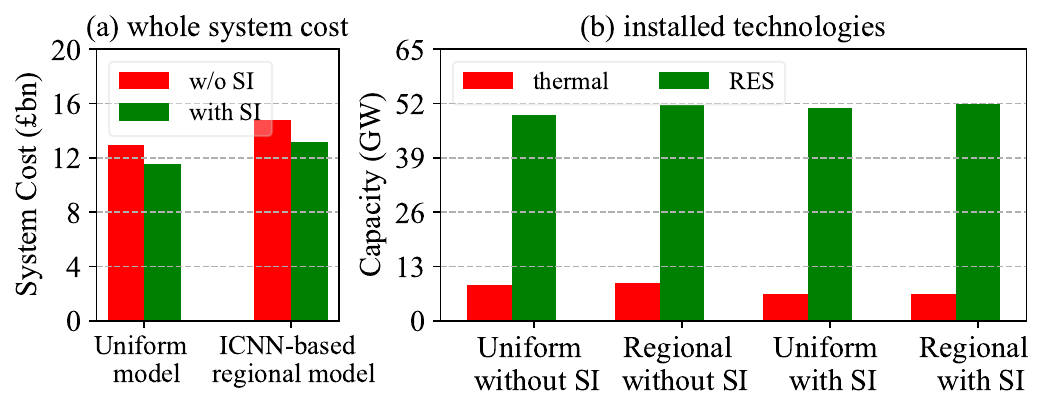}}
\vspace{-1.98em}
\caption{Comparisons between uniform frequency-constrained planning and regional frequency-constrained planning in the modified IEEE 118-bus power system with six regions.}
\label{fig:big_cost_2}
\vspace{-0.58em}
\end{figure}

\vspace{-0.0em}
\label{sec:V.D}
This section serves as a further illustration of the proposed regional frequency-constrained planning approach on scalability through the 14-bus GB power system including several regions (e.g., England, Scotland, etc.) \cite{qiu2023personalized} and a modified IEEE 118-bus system including six regions \cite{shahidehpour2002market}, where the region partition can be achieved by identifying and increasing the impedance between different areas in the power system. In detail, the original three zones provided in \cite{shahidehpour2002market} are further divided into six regions for scalability purposes. Concerning both the uniform frequency model and ICNN-based regional frequency model, the whole system cost and installed capacities of RESs in these two large systems are illustrated in Fig. \ref{fig:big_cost} and Fig. \ref{fig:big_cost_2}, respectively. In addition, the influence of including synthetic inertia (SI) from WTs is analysed in both cases, following the model used in \cite{badesa2022assigning}. Furthermore, relevant sensitivity analysis is conducted to study the influence of carbon targets and RoCoF limits, where results are summarised in Table \ref{table:carbon_rocof}.

\begin{table}[h!]
\vspace{-0.28em}
\footnotesize
\centering
\renewcommand\arraystretch{1.00}
\setlength{\abovecaptionskip}{12pt}
\caption{Sensitivity analysis on the influence of different carbon targets and RoCoF limits concerning the whole system costs.}
\setlength{\tabcolsep}{2.28mm}{
\begin{tabular}{|c|c|c|c|c|}
\toprule
No. & Carbon target & Cost ($\pounds bn$) & RoCoF limit & Cost ($\pounds bn$)  \\ \midrule
1& 100 g/kWh   & 12.8      & 1 Hz/s   & 12.8         \\ \midrule
2& 50 g/kWh  & 13.3      & 0.5 Hz/s   & 13.4         \\ \midrule
3& 25 g/kWh  & 14.0      & 0.125 Hz/s   & 16.5         \\ \bottomrule
\end{tabular}}
\label{table:carbon_rocof}
\vspace{-0.88em}
\end{table}

Based on the 14-bus GB power system, it can be observed from Fig. \ref{fig:big_cost} that considering regional frequency security can lead to higher investment costs than the case with a uniform frequency model. Additionally, considering frequency response from WTs can significantly reduce the whole system costs and increase the accommodation of RESs, compared with the case without wind frequency support. Furthermore, Table \ref{table:carbon_rocof} illustrates that changing the carbon target from 100 g/kWh to 25 g/kWh can drive the increase of the whole system costs, because of the higher requirement for low-carbon technologies (e.g., Gas CCS and RESs). Finally, tightening the RoCoF limit from 1 Hz/s to 0.125 Hz/s can significantly increase the system costs, especially due to the much higher requirement for system inertia.

\begin{figure}[h!]
\vspace{-0.38em}
\centering
{\includegraphics[width=0.49\textwidth]{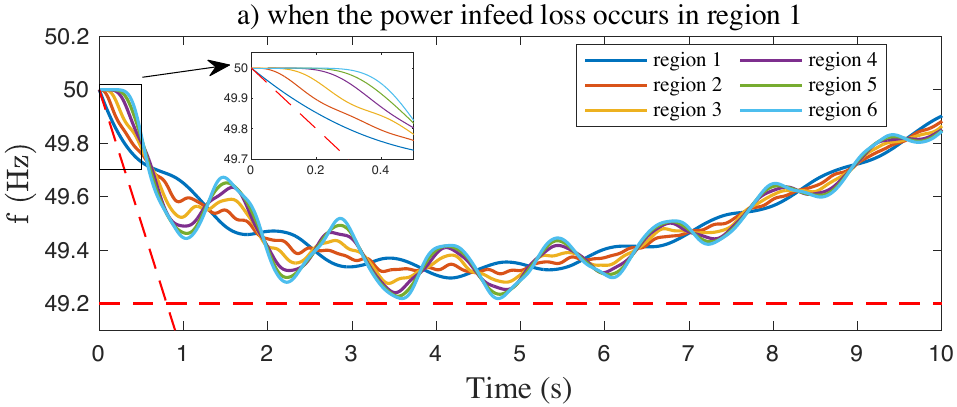}}
{\includegraphics[width=0.49\textwidth]{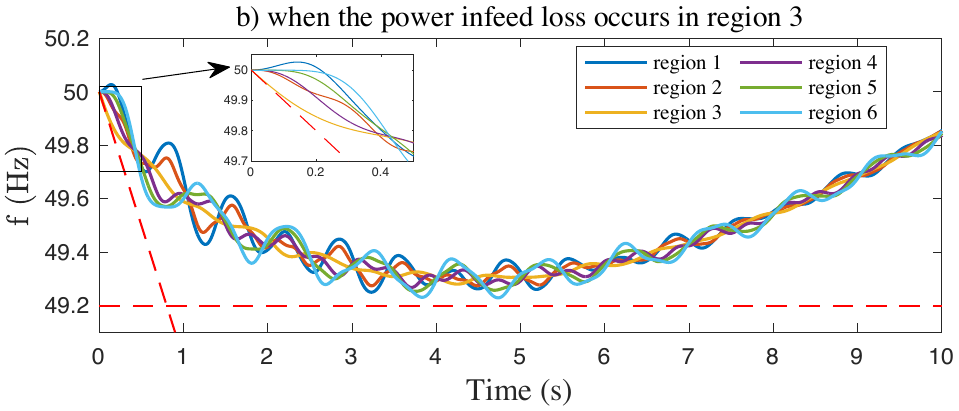}}
\vspace{-1.48em}
\caption{Results of regional frequency deviations from the ICNN-based regional frequency model under a 2 GW power infeed loss.}
\label{fig:matlab_big}
\vspace{-0.68em}
\end{figure}

To further validate the scalability of the proposed regional frequency-constrained planning model for large-scale power systems, a modified IEEE 118-bus system \cite{shahidehpour2002market} is utilised, comprising six regions with distinct frequency deviations. The optimal sizing results obtained under both the uniform frequency model and the proposed ICNN-based regional frequency model are presented in Fig. \ref{fig:big_cost_2}, which align with the conclusions drawn from the 14-bus power system. Additionally, it can be found that considering SI from WTs significantly reduces the investment costs of the ICNN-based regional frequency model, compared with the uniform frequency model. Furthermore, the performance of the enhanced ICNN in capturing inter-area frequency oscillations and ensuring regional frequency security has been verified through dynamic simulations conducted in MATLAB/Simulink. As depicted in Fig. \ref{fig:matlab_big}, the embedded regional frequency constraints effectively guarantee the frequency security of all six regions, i.e., meeting frequency nadir limit 49.2 Hz and RoCoF limit 1 Hz/s, further demonstrating the robustness of the enhanced ICNN in learning regional frequency constraints.

Finally, the computational time of the proposed regional frequency-constrained planning model including AGA and ICNNs is reported in Table \ref{tab:computing}, compared with two benchmark scenarios, i.e., regional frequency-constrained planning with a normal GA and uniform frequency-constrained planning model with AGA. All the case studies were run on Intel Core i7-8700 processor with 16 GB memory.

\begin{table}[h!]
\vspace{-0.18em}
\footnotesize
\centering
\renewcommand\arraystretch{1.00}
\setlength{\abovecaptionskip}{12pt}
\caption{Computational performance of three different methods.}
\setlength{\tabcolsep}{2.78mm}{
\begin{tabular}{ |c|c|c|}
 	  \toprule
        No.
        &\begin{tabular}[c]{@{}c@{}}Method \end{tabular}
         & \begin{tabular}[c]{@{}c@{}} Computation time (hour)\end{tabular}\\ \midrule 
        1 & \begin{tabular}[c]{@{}c@{}}Regional frequency-constrained\\ planning with AGA\end{tabular}  & 5.8 \\ \midrule 
        1 & \begin{tabular}[c]{@{}c@{}}Regional frequency-constrained\\ planning with normal GA\end{tabular}  & 6.5 \\ \midrule 
        2 & \begin{tabular}[c]{@{}c@{}}Uniform frequency-constrained\\ planning with AGA\end{tabular}  &  5.4\\ \bottomrule      
\end{tabular}}
\label{tab:computing}
\vspace{-0.4em}
\end{table}
 
As shown in Table \ref{tab:computing}, the uniform frequency-constrained planning model with AGA achieved the shortest computing time (5.4 hours), as it did not account for inter-area frequency oscillations. Conversely, the regional frequency-constrained planning model with a normal GA had the longest computing time (6.5 hours). After adopting the proposed AGA with sparsity calculation and local search as well as adaptive crossover and mutation probabilities, the computing time of regional frequency-constrained planning has been significantly reduced to 5.8 hours. Given that the primary objective of the proposed model is long-term planning rather than short-term operation, this computing time is considered acceptable.
\vspace{-0.58em}

\section{Conclusions}
\label{sec:VI}
A novel frequency-constrained planning approach is proposed in this paper to solve the optimal sizing problem of power systems capturing regional frequency security. An enhanced ICNN based on the PWI strategy is proposed to extract the relationship between operation conditions and regional frequency security, which is then embedded into the original operation model for regional frequency insurance. The PWI strategy can effectively deal with the issues of non-negative weight matrices and vanishing gradients and improve the fitting ability of ICNNs. An AGA with sparseness and local search is developed to split the planning model into two stages and then solve it iteratively, where a dynamic simulation module is integrated for decision security. In addition, a detailed UC model is developed with different ramping rates, simulating realistic schedules of each generator and leading to accurate inertia calculation. Case studies are conducted on three power systems including a modified IEEE 6-bus system, a 14-bus GB system, and a modified IEEE 118-bus system to evaluate the superiority of the planning approach in ensuring regional frequency security and obtaining cost-effective investment solutions.
\vspace{-0.48em}

\bibliographystyle{IEEEtran}
\bibliography{References.bib}

\begin{IEEEbiography}[{\includegraphics[width=1in,height=1.25in,clip,keepaspectratio]{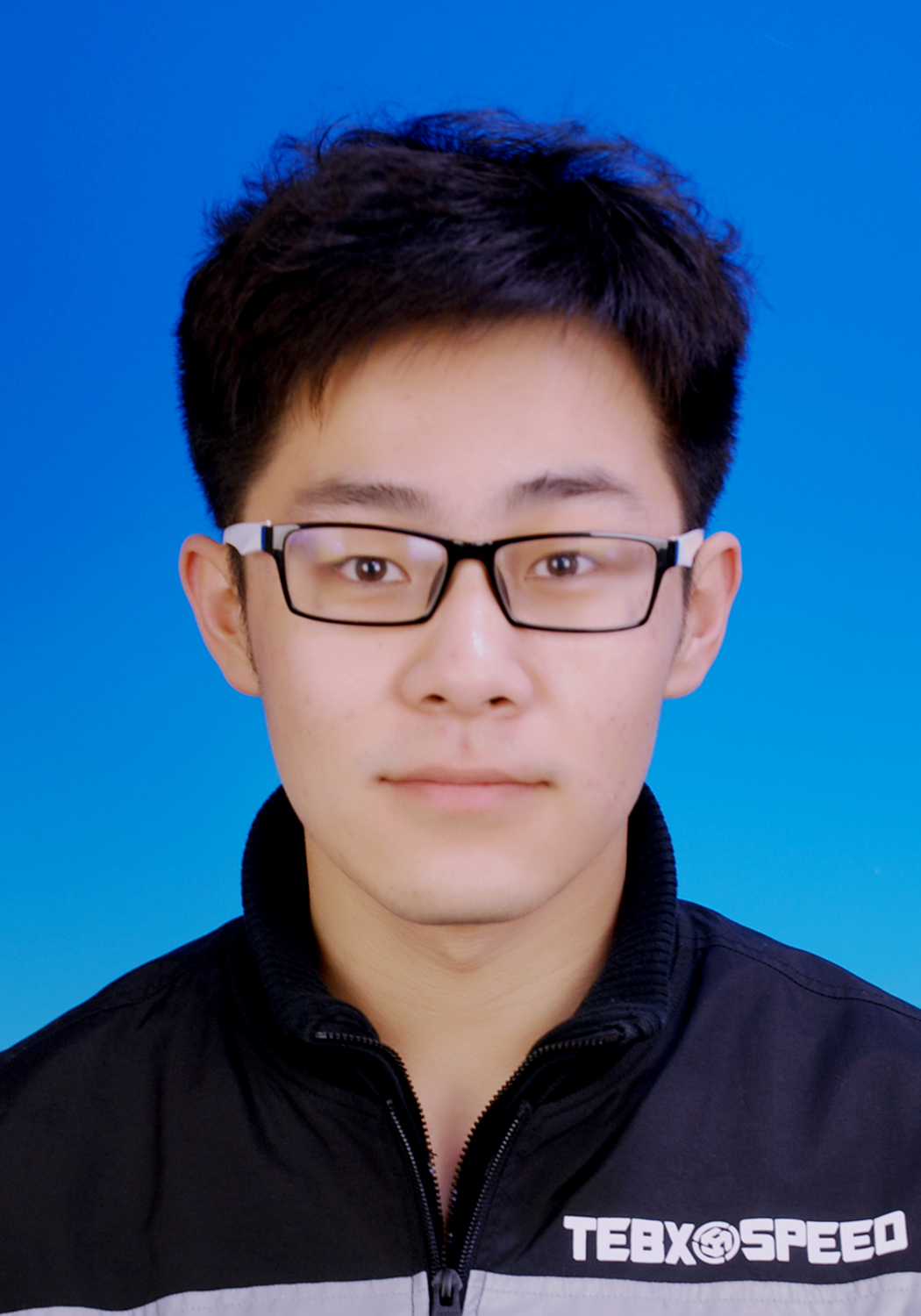}}]{Yi Wang}
(Member, IEEE)
received the B.Eng. degree and the M.Eng. degree from Tianjin University in 2015 and 2018, and the Ph.D. degree from Imperial College London in 2022. He is currently employed as a Research Associate in the Department of Electrical and Electronic Engineering at Imperial College London. His research interests include mathematical programming and learning approaches applied to the planning and operation of networked microgrids, the resilience enhancement of future power systems, frequency-constrained power system optimisation, and multi-energy system integration.
\end{IEEEbiography}

\begin{IEEEbiography}[{\includegraphics[width=1in,height=1.25in,clip,keepaspectratio]{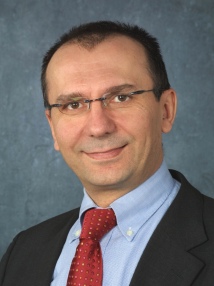}}]{Goran Strbac}
(Member, IEEE) 
is a Professor of Energy Systems at Imperial College London, London, U.K. He led the development of novel advanced analysis approaches and methodologies that have been extensively used to inform industry, governments, and regulatory bodies about the role and value of emerging new technologies and systems in supporting cost effective evolution to smart low carbon future. He is currently the Director of the joint Imperial-Tsinghua Research Centre on Intelligent Power and Energy Systems, Leading Author in IPCC WG 3, Member of the European Technology and Innovation Platform for Smart Networks for the Energy Transition, and Member of the Joint EU Programme in Energy Systems Integration of the European Energy Research Alliance.
\end{IEEEbiography}

\end{document}